\documentclass[twocolumn,prb,floats,aps,amsmath,amssymb,superscriptaddress]{revtex4-1}
\usepackage[]{graphicx}
\usepackage{color}
\usepackage{bm} 
\usepackage{prettyref}
\usepackage{float}
\usepackage{natbib}

\usepackage{array}

\begin{document}
	
	\title{Coupled commensurate charge density wave and lattice distortion in Na$_{2}$Ti$_{2}Pn_{2}$O ($Pn =$ As, Sb) determined by X-ray diffraction and angle-resolved photoemission spectroscopy}
	\date{\today}
	\author{N. R. Davies}
	\author{R. D. Johnson}
	\author{A. J. Princep}
	\author{L. A. Gannon}
	\affiliation{Department of Physics, University of Oxford, Clarendon Laboratory, Oxford, OX1 3PU, U.K.}
	\author{J.-Z. Ma}
	\author{T. Qian}
	\affiliation{Beijing National Laboratory for Condensed Matter Physics \& Institute of Physics, Chinese Academy of Science, Beijing 100190, China}
	\author{P. Richard}
	\affiliation{Beijing National Laboratory for Condensed Matter Physics \& Institute of Physics, Chinese Academy of Science, Beijing 100190, China}
	\affiliation{Collaborative Innovation Center of Quantum Matter, Beijing, China}
	\author{H. Li}
	\affiliation{Beijing National Laboratory for Condensed Matter Physics \& Institute of Physics, Chinese Academy of Science, Beijing 100190, China}
	\author{H. Nowell}
	\affiliation{Diamond Light Source, Ltd., Harwell Science and Innovation Campus, Didcot, Oxfordshire OX11 0DE, U.K.}
	\author{P. J. Baker}
	\affiliation{ISIS Facility, STFC Rutherford Appleton Laboratory, Chilton, Didcot OX11 0QX, U.K.}
	\author{Y. G. Shi}
	\affiliation{Beijing National Laboratory for Condensed Matter Physics \& Institute of Physics, Chinese Academy of Science, Beijing 100190, China}
	\author{H. Ding}
	\affiliation{Beijing National Laboratory for Condensed Matter Physics \& Institute of Physics, Chinese Academy of Science, Beijing 100190, China}
	\affiliation{Collaborative Innovation Center of Quantum Matter, Beijing, China}
	\author{J. Luo}
	\affiliation{Beijing National Laboratory for Condensed Matter Physics \& Institute of Physics, Chinese Academy of Science, Beijing 100190, China}
	\author{Y. F. Guo}
	\email{guoyf@shanghaitech.edu.cn}
	\affiliation{Department of Physics, University of Oxford, Clarendon Laboratory, Oxford, OX1 3PU, U.K.}
	\affiliation{School of Physical Science and Technology, ShanghaiTech University, 319 Yueyang Road, Shanghai 200031, China}
	\author{A. T. Boothroyd}
	\email{a.boothroyd@physics.ox.ac.uk}
	\affiliation{Department of Physics, University of Oxford, Clarendon Laboratory, Oxford, OX1 3PU, U.K.}

\begin{abstract}
	We report single crystal X-ray diffraction measurements on Na$_2$Ti$_{2}Pn_{2}$O ($Pn$ = As, Sb) which reveal a charge superstructure that appears below the density wave transitions previously observed in bulk data. From symmetry-constrained structure refinements we establish that the associated distortion mode can be described by two propagation vectors, ${\bf q}_{1} = (1/2, 0, l)$ and ${\bf q}_{2} = (0, 1/2, l)$, with $l=0$ (Sb) or $l = 1/2$ (As), and primarily involves in-plane displacements of the Ti atoms perpendicular to the Ti--O bonds. We also present angle resolved photoemission spectroscopy (ARPES) measurements, which show band folding and back bending consistent with a density wave with the same wave vectors ${\bf q}_{1}$ and ${\bf q}_{2}$ associated with fermi surface nesting, and muon-spin relaxation data, which show no indication of spin density wave order. The results provide direct evidence for phonon-assisted charge density wave order in Na$_2$Ti$_{2}Pn_{2}$O and fully characterise a proximate ordered phase that could compete with superconductivity in doped BaTi$_{2}$Sb$_{2}$O.
\end{abstract}

\maketitle


\begin{figure}[t]
	\centering
	\includegraphics[width=0.9\columnwidth,clip]{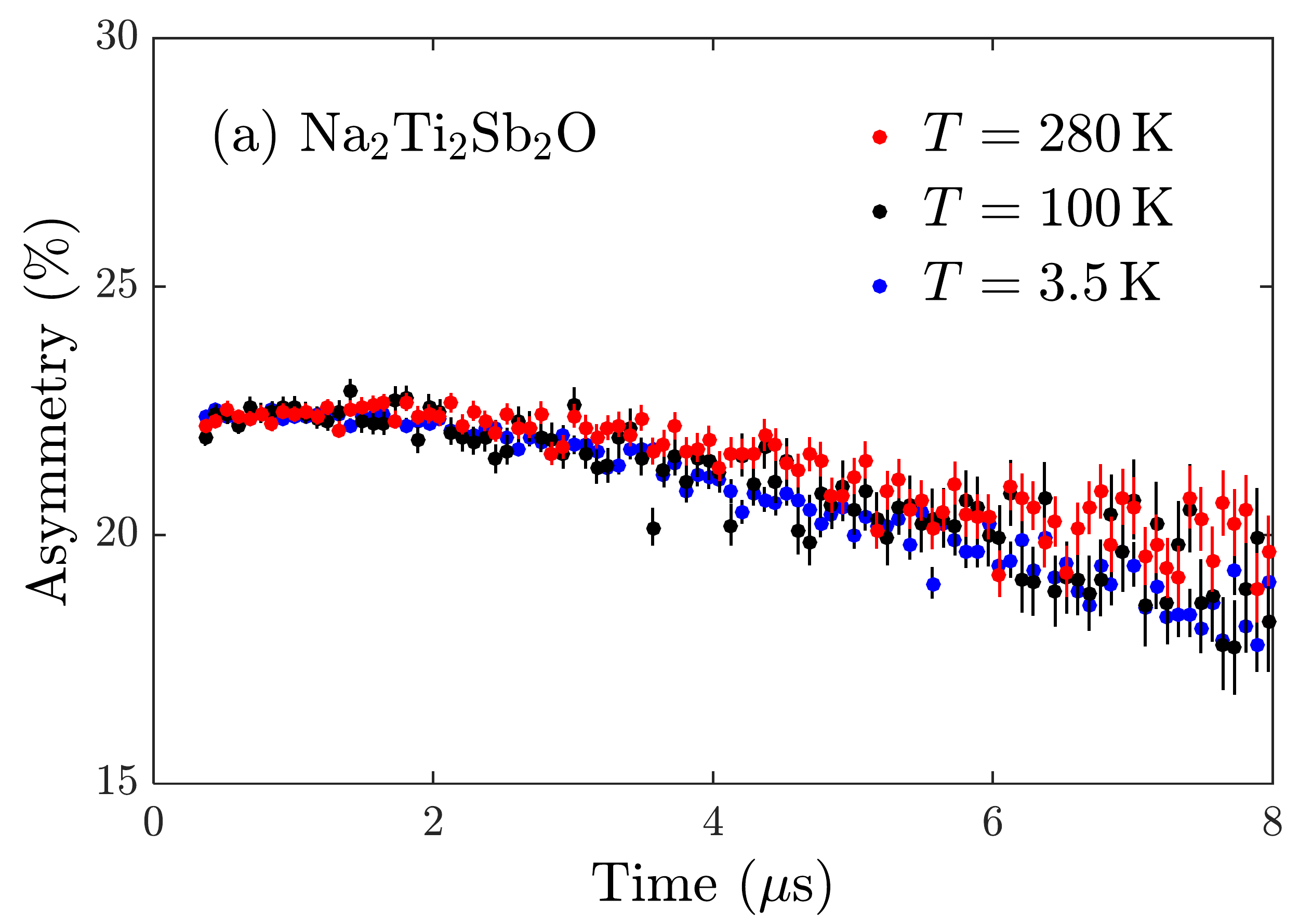}
	\includegraphics[width=0.9\columnwidth,clip]{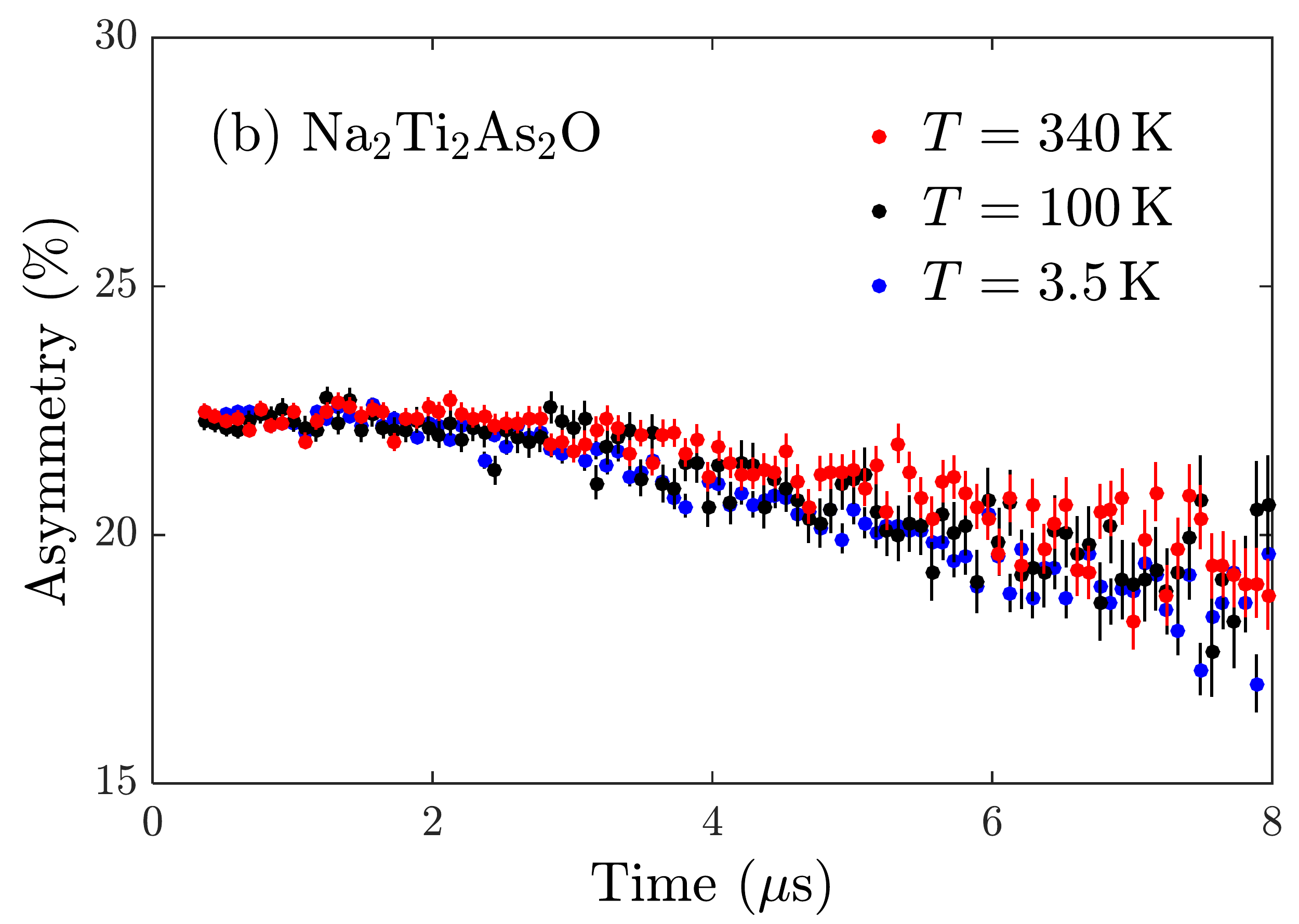}
	\caption{Temperature dependence of the $\mu$SR asymmetry for (a) Na$_{2}$Ti$_{2}$Sb$_{2}$O and (b) Na$_{2}$Ti$_{2}$As$_{2}$O. There is no evidence in the curves for any relaxation due to magnetic order or magnetic fluctuations.}
	\label{FigS1}
\end{figure}

	\begin{figure*}[t]
		\centering
		\includegraphics[width=0.9\textwidth]{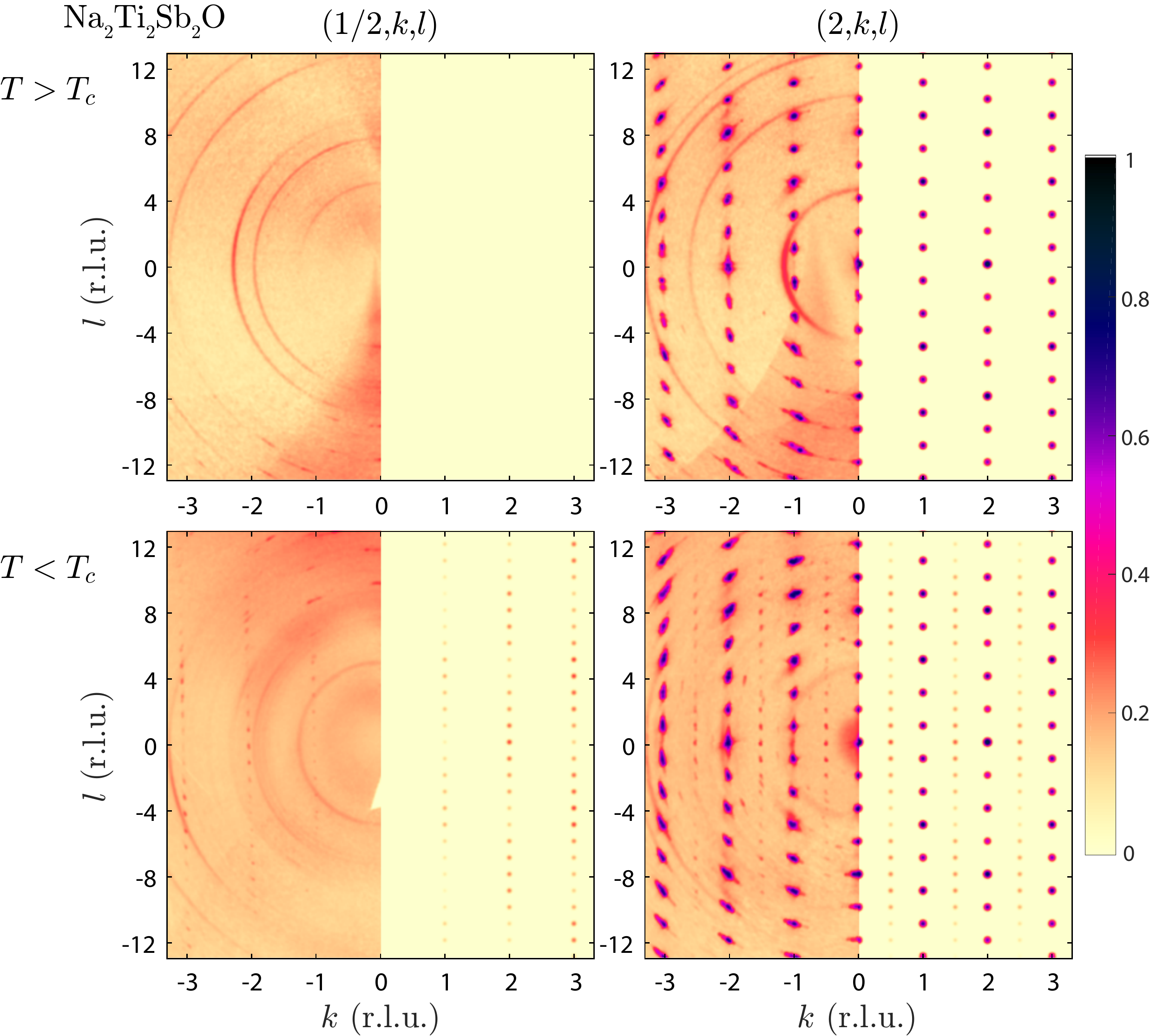}
		\caption{Experimental single crystal X-ray diffraction maps (left half-panels) and simulations (right half-panels) in the $(1/2,k,l)$ and $(2,k,l)$ reciprocal space planes of Na$_{2}$Ti$_{2}$Sb$_{2}$O measured at temperatures above (300\,K) and below ($\sim $100\,K) the structural distortion temperature $T_{\rm DW}$. Intensities are plotted on a log scale.}
		\label{XRD_Planes_Fig_Sb}
	\end{figure*}
	
	\begin{figure*}[t]
		\centering
		\includegraphics[width=0.9\textwidth]{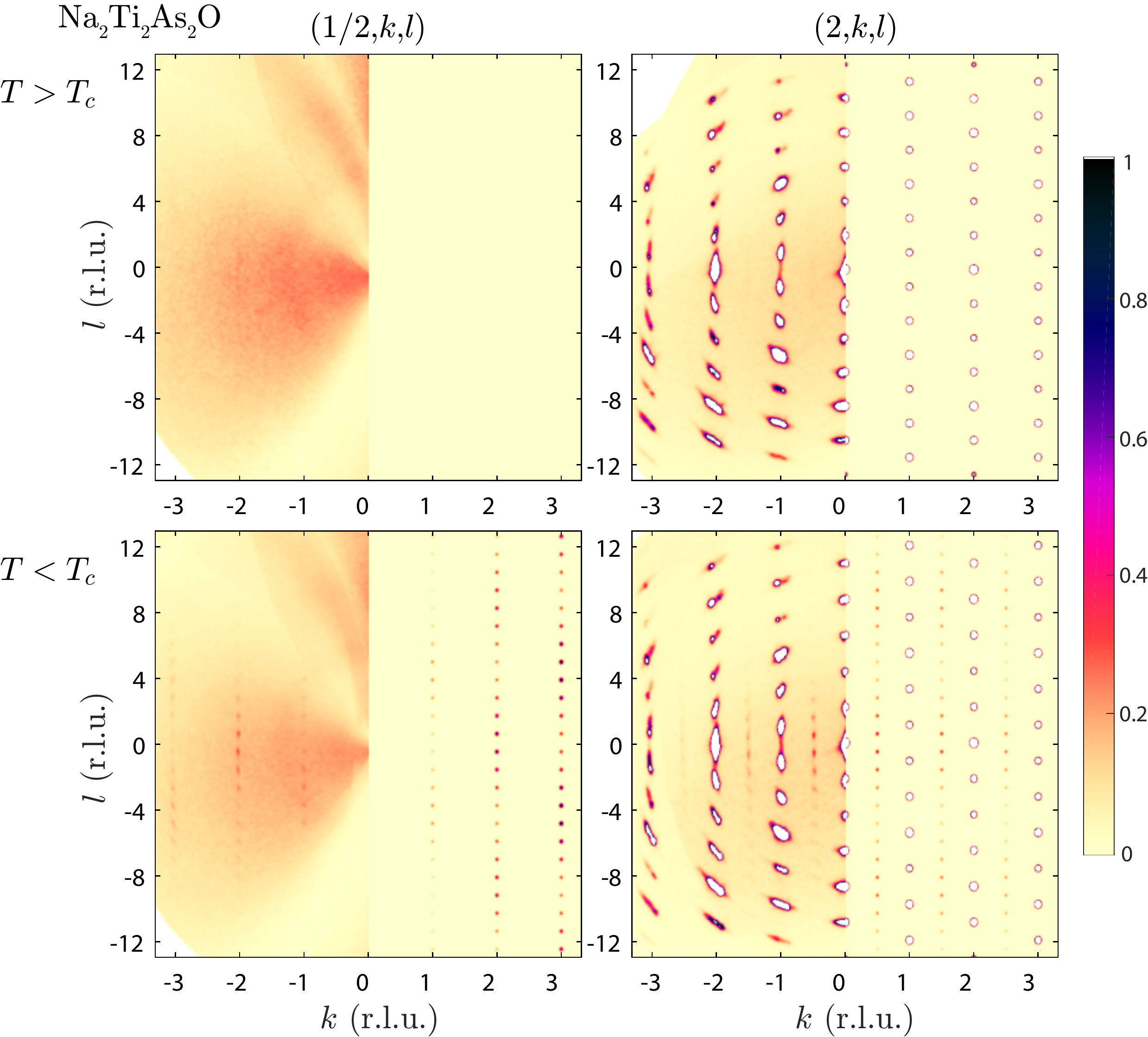}
		\caption{Experimental single crystal X-ray diffraction maps (left half-panels) and simulations (right half-panels) in the $(1/2,k,l)$ and $(2,k,l)$ reciprocal space planes of Na$_{2}$Ti$_{2}$As$_{2}$O measured at temperatures above (330\,K) and below (230\,K) the structural distortion temperature $T_{\rm DW}$. Intensities are plotted on a log scale.}
		\label{XRD_Planes_Fig_As}
	\end{figure*}
		
\begin{figure*}
	\centering
	\includegraphics[width=\textwidth]{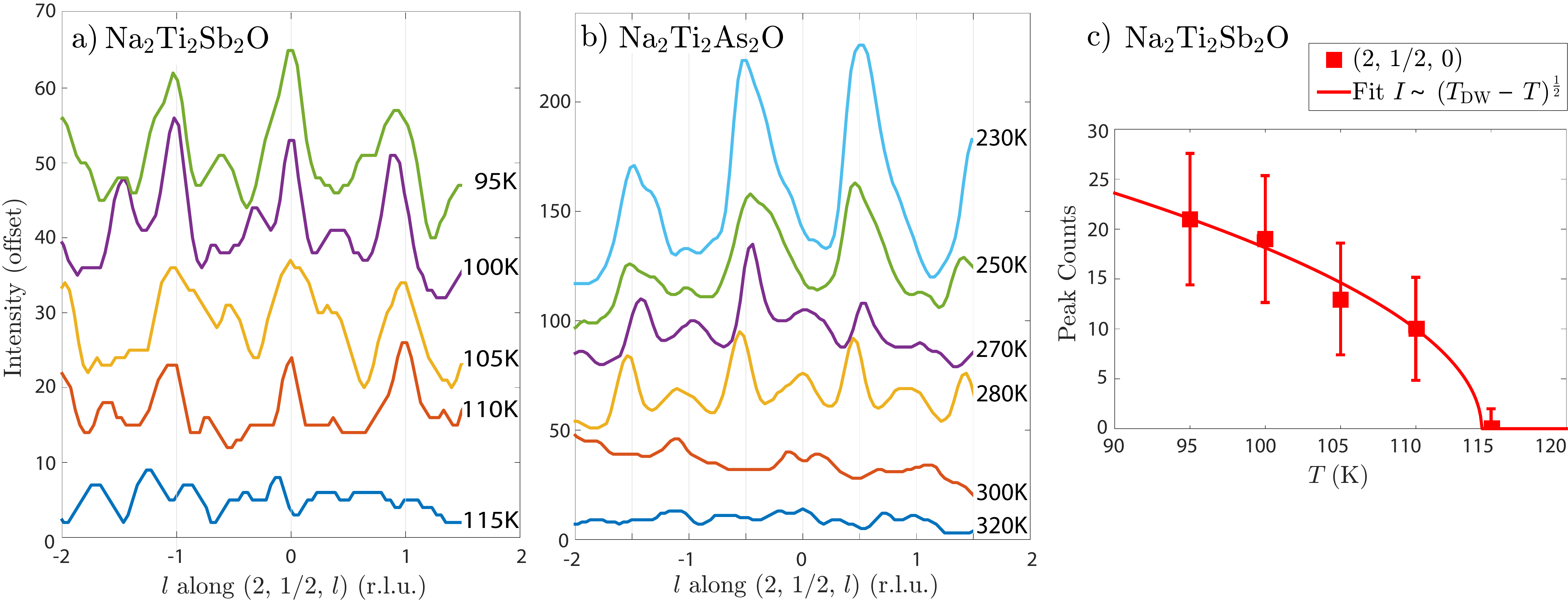}
	\caption{Measured X-ray intensity as a function of temperature close to $T_{\rm DW}$ along the cut $(2,1/2,l)$ for (a) Na$_{2}$Ti$_{2}$Sb$_{2}$O and (b) Na$_{2}$Ti$_{2}$As$_{2}$O. Successive cuts have been offset vertically for clarity. (c) Intensity at the maximum of the $(2,1/2,0)$ peak of Na$_{2}$Ti$_{2}$Sb$_{2}$O. The red line is a best fit of the form $I \sim {(T_{\rm DW}-T)}^{1/2}$, intended as a guide to the eye.}
	\label{XRD_Cuts_fig}
\end{figure*}

\section{Introduction} 

The appearance of superconductivity in the neighbourhood of other symmetry-breaking ground states is a prominent characteristic of many unconventional superconductors.\cite{Monthoux2007} In some theoretical scenarios strong fluctuations of the proximate order parameter can contribute to the pairing interaction, so it is important to establish the identity of the competing orders and to determine the relevant broken symmetries.

The proximity of superconductivity to spin density wave (SDW) order is known in a range of different systems, and is a particularly robust feature of the iron-based superconductors. In these materials, the SDW is believed to be assisted by partial nesting of quasi-two-dimensional electron and hole Fermi surface pockets,\cite{Mazin2008, Scalapino2012} and the strong spin fluctuations that remain after suppression of the SDW are widely thought to play a central role in mediating unconventional superconductivity.\cite{Chubukov2012} Interplay between superconductivity and charge density wave (CDW) formation is less often encountered but is known in several transition-metal chalcogenides and oxides,\cite{Wilson1975,Morosan2006,Yokoya2005,Mattheiss1988}
especially the layered copper oxides where CDW order is found to compete with superconductivity in several hole-doped materials near 1/8 doping. \cite{Tranquada1995,fujita2004,Croft2014,daSilvaNeto2014,Wu2011,Ghiringhelli2012,Chang2012} Theoretical work suggests that charge fluctuations near the onset of CDW order may play an important role in the superconducting pairing mechanism.\cite{Wang2015}

The tetragonal layered titanium oxypnictides \cite{Lorenz2014} $A$Ti$_{2}Pn_{2}$O ($A =$ Ba, Na$_{2}$, (SrF)$_{2}$, (SmO)$_{2}$; $Pn =$ As, Sb, Bi) share structural similarities with the layered copper oxide and Fe-based families of superconductors and display anomalies in magnetic, transport and thermal data at various temperatures up to 400\,K which have been interpreted to indicate density wave (DW) transitions.\cite{Adam1990,Axtell1997,Ozawa2000,Ozawa2001,Liu2009,Liu2010,Wang2010,Yajima2012} The recent discovery of superconductivity in BaTi$_{2}$Sb$_{2}$O at $T_{\rm c}\simeq 1.2$\,K in the vicinity of a possible DW phase appearing at $T_{\rm DW} \simeq 55$\,K has renewed interest in this family of materials.\cite{Yajima2012,Doan2012} Substitution of Na for Ba increases $T_{\rm c}$ up to 5.5\,K with concomitant gradual suppression of the DW transition.\cite{Doan2012} A similar anticorrelation between $T_{\rm c}$ and $T_{\rm DW}$ has been observed with other substitutions indicating competition between the superconducting and DW phases, although there remain large regions of the phase diagrams in which they appear to coexist.\cite{Yajima2013,Zhai2013,Nakano2013,Pachmayr2014,vonRohr2014}

Many attempts have been made to identify the DW phase in the titanium oxypnictides. Electronic structure models predict a highly anisotropic Fermi surface with box-like electron and hole pockets that are quite well nested and therefore susceptible to either SDW or CDW instabilities \cite{Pickett1998,deBiani1998,Singh2012,Wang2013,Yan2013,Suetin2013,Yu2014}. The calculations indicate that strong electron--phonon coupling could induce a superlattice distortion and accompanying CDW.\cite{Subedi2013,Nakano2016} The predicted Fermi surface agrees well with data from angle-resolved photoemission spectroscopy (ARPES), which also shows evidence for gap formation at the DW transitions on parts of the Fermi surface.\cite{Tan2015,Song2016} An initial neutron powder diffraction study revealed anomalies in the lattice parameters of Na$_2$Ti$_2$Sb$_2$O at $T_{\rm DW} \simeq 115$\,K.\cite{Ozawa2000} Subsequent electron and neutron diffraction studies of $A$Ti$_{2}Pn_{2}$O compounds have not found any bulk magnetic or charge superstructure associated with a SDW or CDW ordering,\cite{Nozaki2013,Frandsen2014} although there is evidence for tetragonal symmetry breaking in BaTi$_{2}$(As,Sb)$_{2}$O from neutron diffraction \cite{Frandsen2014} and nuclear quadrupole resonance (NQR),\cite{Kitagawa2013} and in Na$_2$Ti$_{2}$As$_{2}$O from Raman scattering \cite{Chen2016}. Muon-spin relaxation ($\mu$SR) studies of Na-doped BaTi$_{2}$(As$_{1-x}$Sb$_{x}$)$_{2}$O did not detect any static magnetic moments, leading to the conclusion that the DW phase is most likely a CDW,\cite{Nozaki2013,vonRohr2013} and the NQR study concluded that only a commensurate CDW could explain the data.\cite{Kitagawa2013} The lack of any evidence for an accompanying CDW superstructure, however, has recently prompted proposals that the DW phase is either some form of intra- unit cell orbital nematic state,\cite{Frandsen2014,Nakaoka2016} or is characterized by an orbital polarization that takes place without any lowering of symmetry.\cite{Kim2015}

In this work we studied Na$_2$Ti$_{2}$Sb$_{2}$O and Na$_2$Ti$_{2}$As$_{2}$O by single crystal X-ray diffraction, $\mu$SR and ARPES. We identify the DW phase as a commensurate two-{\bf q} CDW through the accompanying structural distortion. The distortion results in a $2 \times 2$ superstructure within the Ti$_2$O layers with transverse displacements of the Ti atoms and much smaller shifts in the Sb/As and Na positions. The results explain the observation of band folding consistent with the CDW propagation vectors in our ARPES data, and imply a strong electron--phonon coupling which could be responsible for superconductivity in doped BaTi$_{2}$Sb$_{2}$O.

\section{Experimental details}

Single crystals and powder samples of Na$_{2}$Ti$_{2}$Sb$_{2}$O and Na$_{2}$Ti$_{2}$As$_{2}$O were prepared by the method described by Shi {\it et al.},\cite{Shi2013} with magnetic and electrical measurements confirming that the DW phase transitions in Na$_{2}$Ti$_{2}Pn_{2}$O occur at $T_{\rm DW}\simeq 115$\,K and $T_{\rm DW}\simeq 320$\,K for $Pn =$ Sb and As, respectively.\cite{Shi2013}

$\mu$SR measurements were carried out on the EMU spectrometer at the ISIS Pulsed Muon Facility using powder samples of Na$_{2}$Ti$_{2}Pn_{2}$O ($Pn = $ Sb, As) packed inside 25$\mu$m silver foil packets mounted on a silver backing plate and measured at temperatures above and below $T_{\rm DW}$ using a closed-cycle refrigerator. 


Temperature dependent X-ray diffraction measurements were performed on single crystal samples of Na$_{2}$Ti$_{2}Pn_{2}$O ($Pn = $ Sb, As) using a Mo-source Oxford Diffraction Supernova diffractometer equipped with a liquid nitrogen flow cryostat. The diffraction data presented here on Na$_{2}$Ti$_{2}$As$_{2}$O were collected from a large single crystal of approximate dimensions $2\times 0.5 \times 1$\,mm$^3$, while data taken on Na$_{2}$Ti$_{2}$Sb$_{2}$O were from a much smaller sample $0.5\times0.5\times0.1$\,mm$^{3}$. Additionally, synchrotron X-ray diffraction patterns were recorded from a single-crystal of Na$_{2}$Ti$_{2}$Sb$_{2}$O at high ($T=$ 300\,K $>T_{\rm DW}$) and low ($T\sim$ 100\,K $<T_{\rm DW}$) temperatures on the I19 beamline \cite{Nowell2012} at the Diamond Light Source. The experiment was conducted in the experimental hutch EH1 using an incident monochromatic beam at the Zr edge with a wavelength of 0.6889\,\AA, and a 4-circle diffractometer with a CCD detector. The Helix device installed on the beamline was used to cool the sample below its transition temperature using a flow of helium. These materials are air-sensitive and were coated with vacuum grease to prevent decomposition during diffraction measurements. Data from both compounds show some disorder along the $c$-axis likely due to random stacking faults which may be anticipated in systems with weak inter-layer chemical bonding.

All Bragg peaks in the highest temperature datasets for both compositions could be indexed in the $I4/mmm$ space group, and we shall henceforth refer Bragg reflections and reciprocal lattice coordinates to the $I4/mmm$ conventional unit cell with lattice parameters $a = 4.16$\,\AA, $c = 16.6$\,\AA\ ($Pn = $ Sb) \cite{Ozawa2000} or $a = 4.08$\,\AA, $c = 15.3$\,\AA\ ($Pn = $ As) \cite{Ozawa2001} at room temperature.

ARPES measurements were performed on single crystals of Na$_2$Ti$_2$Sb$_2$O at beamline PGM (Plane Grating Monochromator) of the Synchrotron Radiation Center (Wisconsin), as well as at the beamline SIS (Surface and Interface Spectroscopy) of the Swiss Light Source at Paul Scherrer Institute (PSI), both equipped with a Scienta R4000 analyzer. The energy and angular resolutions were set at 15-30 meV and 0.2$^{\circ}$, respectively. The crystals were cleaved \textit{in situ} and measured in the 24 to 150\,K temperature range in a vacuum better than $8\times 10^{-11}$ Torr. The ARPES data were recorded using $s$ polarized light.

\section{Results and analysis}

\subsection{X-ray diffraction}

Figure~\ref{FigS1} shows the temperature variation of the measured zero-field $\mu$SR asymmetry after muon implantation. The asymmetry shows no characteristic features associated with magnetic order or magnetic fluctuations in either sample and has a very small relaxation rate which is virtually independent of temperature. Furthermore, there was no reduction in the asymmetry of the muon spin precession in a weak transverse field. This observed $\mu$SR behaviour is most likely dominated by nuclear relaxation from Na and Sb/As nuclei with the very small temperature variation plausibly caused by muon diffusion at high temperatures. These results strongly indicate that the DWs in Na$_{2}$Ti$_{2}Pn_{2}$O are most likely CDWs rather than SDWs.

Figures~\ref{XRD_Planes_Fig_Sb} and \ref{XRD_Planes_Fig_As} present X-ray diffraction intensity maps for Na$_{2}$Ti$_{2}$Sb$_{2}$O and Na$_{2}$Ti$_{2}$As$_{2}$O respectively in the $(1/2,k,l)$ and $(2,k,l)$ reciprocal space planes recorded above and below $T_{\rm DW}$. For $Pn = $ Sb the data reveal weak superstructure reflections at positions with indices $h+k =$ half an odd integer, $l =$ integer which vanish above $T_{\rm DW}$, while for $Pn =$As a similar superstructure is observed except the peaks are shifted to half-integer positions along $l$. This superstructure is our key experimental observation.

 The precise $l$ positions of the superstructure and its evolution as a function of temperature close to $T_{\rm DW}$ is best seen by taking cuts through the data from the laboratory X-ray diffractometer, as shown in Fig.~\ref{XRD_Cuts_fig}. Figure~\ref{XRD_Cuts_fig}(a) shows the superstructure peaks in Na$_{2}$Ti$_{2}$Sb$_{2}$O in cuts along $(2,1/2,l)$, $-2 \le l \le 2$ at several temperatures between 95\,K and 115\,K, Fig.~\ref{XRD_Cuts_fig}(b) shows a similar plot for Na$_{2}$Ti$_{2}$As$_{2}$O at temperatures between 230\,K and 320\,K and Fig.~\ref{XRD_Cuts_fig}(c) follows the $(2,1/2,0)$ peak maximum intensity as a function of temperature for Na$_{2}$Ti$_{2}$Sb$_{2}$O. The superstructure peaks were detected for temperatures below 115\,K ($Pn = $ Sb) and 300\,K ($Pn = $ As), consistent with the previously measured anomalies in bulk data at $T_{\rm DW}\simeq 115$\,K ($Pn = $ Sb) and 320\,K ($Pn = $ As). Mosaic broadening made it impossible to extract a reliable quantitative temperature dependence for the superstructure peaks for $Pn = $ As.

The superstructures in both materials can be indexed (after averaging over an equal population of equivalent domains) using either a two-$\bf{q}$ distortion with commensurate propagation vectors ${\bf q}_{1}$ and ${\bf q}_2$, or a single-$\bf{q}$ distortion with either of ${\bf q}_{1}$ or ${\bf q}_2$, where ${\bf q}_{1} = (1/2, 0, l)$ and ${\bf q}_{2} = (0, 1/2, l)$, with $l=0$ (Sb) or $l = 1/2$ (As).
Symmetry analysis for these propagation vectors yields a large number of possible distortion modes. We were able to constrain this number significantly by testing each against the qualitative features of our data shown in Figs.~\ref{XRD_Planes_Fig_Sb} and \ref{XRD_Planes_Fig_As}, including the periodic modulation in superstructure peak intensity along $(0,0,l)$ with a repeat of approximately $8c^*$ ($c^* = 2\pi/c$), preservation of absences at $h+k+l = $ odd integer positions for $Pn = $ Sb (i.e.~the absences of the $I4/mmm$ structure) and the fact that in the $(1/2,k,l)$ plane the modulation shifts by half a period along $(0,0,l)$ between odd and even $k$ positions. For each material this eliminated all except one two-$\bf{q}$ mode and two single-$\bf{q}$ modes. In both cases the two-$\bf{q}$ mode involves Ti displacements perpendicular to the local O--Ti--O bond, whereas both single-$\bf{q}$ modes involve a mixture of displacements parallel and perpendicular to O--Ti--O bonds. As transverse displacements are expected to be energetically more favourable than longitudinal ones the two-$\bf{q}$ mode is the most likely distortion, so for each material we performed least-squares refinements of the two-$\bf{q}$ mode against the integrated intensities of the best resolved superstructure peaks at the lowest temperatures. Further details of the symmetry analysis can be found in the Appendix.

\begin{figure}
			\centering
			\includegraphics[width=\columnwidth]{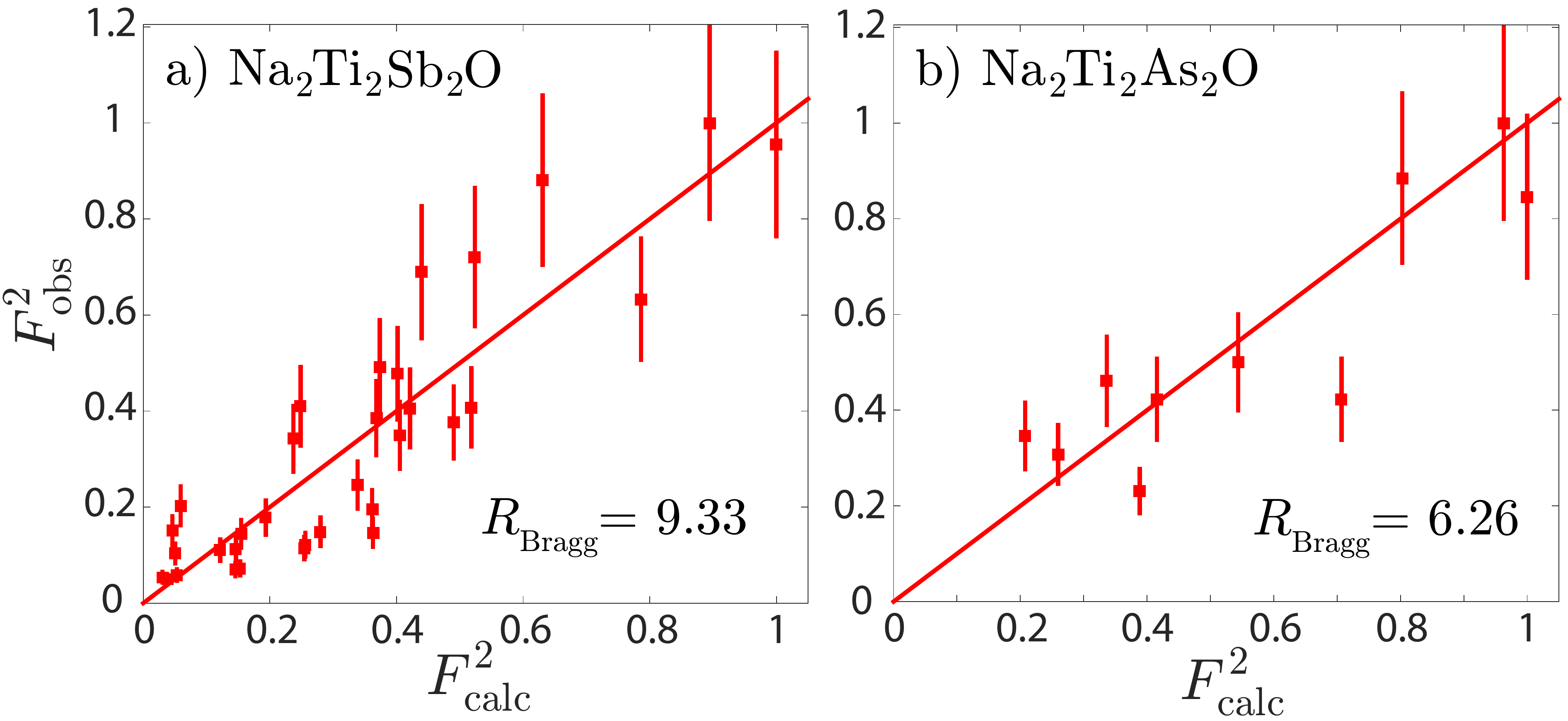}
			\caption{Comparison of observed and calculated modulus-squared structure factors for the superstructure peaks of Na$_{2}$Ti$_{2}Pn_{2}$O. The calculated values are obtained from the distortion models presented in the main text. Error bars represent estimated random errors, but an additional systematic error due to absorption affects some of the points. The $R_{\rm Bragg}$ values are high, but typical of superstructure refinements.}
			\label{F2obsF2cal_Fig}
\end{figure}

The X-ray diffraction intensities calculated from the model are shown in Fig.~\ref{XRD_Planes_Fig_Sb} for $Pn = $ Sb and in Fig.~\ref{XRD_Planes_Fig_As} for $Pn = $ As (right half-panels), and a plot of the observed versus calculated squared structure factors in both cases is given in Fig.~\ref{F2obsF2cal_Fig}. For both materials there is reasonable agreement between the observed and calculated superstructure intensities. In particular, the intensity modulation along $(0,0,l)$ is well reproduced.

\begin{figure*}
	\centering
	\includegraphics[width=0.9\linewidth]{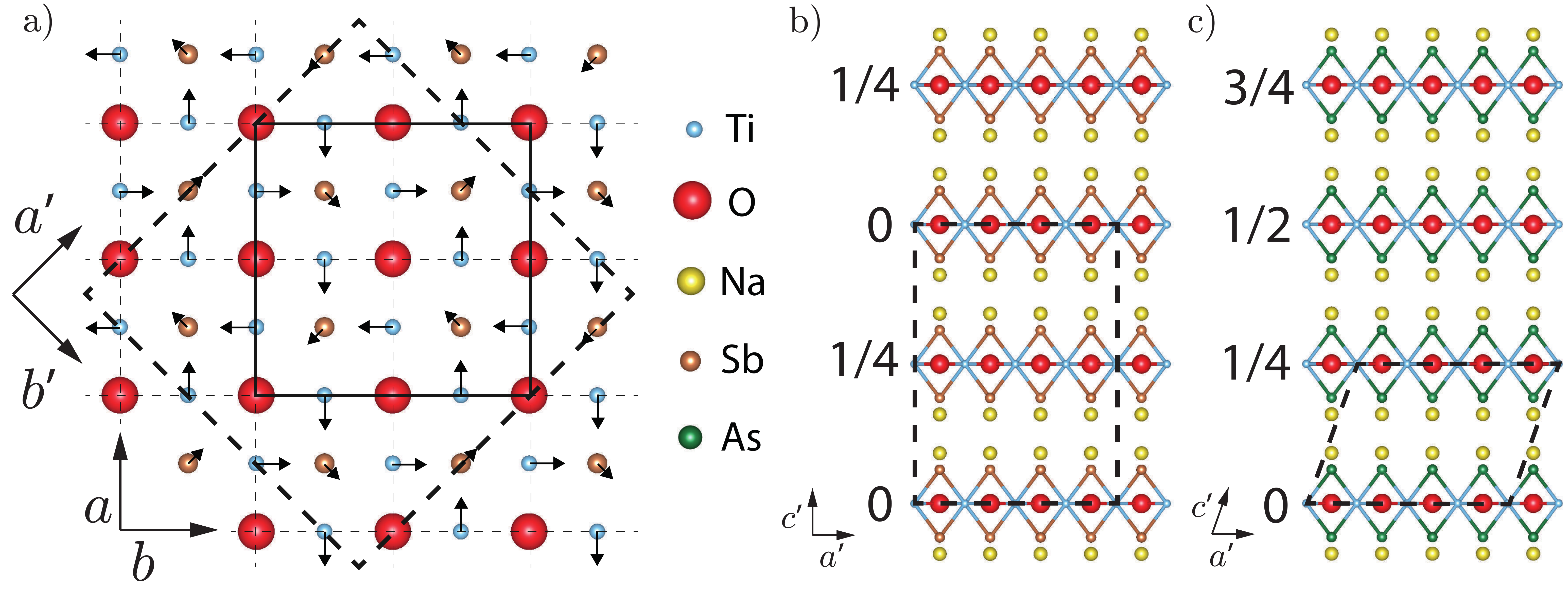}
	\caption{Proposed distortion model for Na$_{2}$Ti$_{2}Pn_{2}$O. (a) $ab$-plane distortion pattern of the Ti$_2Pn_2$O units. Thin dashed lines show the $I4/mmm$ unit cell boundaries, thick dashed lines the low temperature $Cmcm$ unit cell boundary and solid lines the unit cell of the Ti$_2Pn_2$O layer distortion pattern. (b) and (c) show how these layers stack along $c$, with the numbers beside each Ti$_2Pn_2$O unit representing the fractional translation along the orthorhombic $a'$-axis that relates the distortion pattern in that layer to the distortion pattern in the base layer. Black dashed lines mark the unit cells of the distorted structures.}
	\label{Dist_Xtal_Struct_Fig}
\end{figure*}

Figure~\ref{Dist_Xtal_Struct_Fig} shows the distortion models for both materials and the relation between them.
The in-plane displacement pattern of a layer of Ti$_2Pn_2$O units depicted in Fig.~\ref{Dist_Xtal_Struct_Fig}(a), which is the same for both compounds, is a $2 \times 2$ superstructure with the most significant distortion being an in-plane shift of all the Ti atoms perpendicular to the Ti--O nearest-neighbor bonds. The magnitude of the shift is found to be about $0.14 \pm 0.03$\,\AA\ for $Pn =$ Sb and $0.10 \pm 0.03$\,\AA\ for $Pn =$ As. The $Pn$ sites immediately above and below each Ti$_2$O plaquette undergo a smaller in-plane distortion ($\simeq 0.012$\,\AA\ in both cases). Best agreement with experiment for $Pn =$ Sb is obtained when the Na layers are allowed to distort slightly (by $\simeq 0.025$\,\AA) such that every Na atom moves parallel to the nearest $Pn$ immediately above/below it along the $c$ axis (see Appendix). We note, however, that the fit was not very sensitive to the size of the Na distortion.

In the distorted phase the isolated Ti$_2Pn_2$O layer has 4-fold axes through half of the O sites, which probably explains why the $a$ and $b$ lattice parameters remain virtually the same.\cite{Ozawa2000} The 4-fold axes do not coincide between layers upon stacking in the $c$ direction, so the three-dimensional superstructures do not have 4-fold symmetry. 
For $Pn =$ Sb, the superstructure within one Ti$_2$Sb$_2$O layer displaces first by $\bf t$ and then by $-{\bf t}$, where ${\bf t} = (1/2, 1/2, 1/2)$ in $I4/mmm$. The resulting structure is orthorhombic with space group $Cmcm$ and lattice vectors ${\bf a'} = 2({\bf a} + {\bf b})$, ${\bf b'} = 2({\bf a} - {\bf b})$ and ${\bf c'} = {\bf c}$, see Figs.~\ref{Dist_Xtal_Struct_Fig}(a) and (b). In the case of $Pn =$ As, each layer is displaced by ${\bf t}$ relative to the one below resulting in a monoclinic structure with space group $C2/m$ and lattice vectors ${\bf a'} = 2({\bf a} + {\bf b})$, ${\bf b'} = 2({\bf a} - {\bf b})$ and ${\bf c'} = \frac{1}{2}({\bf a} + {\bf b} + {\bf c})$, Fig.~\ref{Dist_Xtal_Struct_Fig}(c).  The distortion in the As compound has period $2c$ along the $c$ axis which is why the superlattice peaks appear at $l =$ half-odd integer positions (half-even integer reflections are absent because the distortion of a Ti$_2$As$_2$O unit undergoes a phase shift of $\pi$ upon translation by $c$ along the $c$ axis.)
The diffraction maps shown in Fig.~\ref{XRD_Planes_Fig_Sb} (for $Pn =$ Sb) and in Fig.~\ref{XRD_Planes_Fig_As} (for $Pn =$ As) are averaged over equal populations of equivalent domains. 



The superstructures found here have some features in common with the distortion mode predicted from density functional theory (DFT) by Subedi \cite{Subedi2013} for BaTi$_{2}$Sb$_{2}$O. Both involve transverse displacements of the Ti atoms relative to the O--Ti--O bond, and the calculated shift of $\simeq 0.14$\,\AA\ is very close to that obtained from our model. However, the predicted distortion mode has a $\sqrt{2}a\times \sqrt{2}a$ in-plane unit cell which has a smaller area and different propagation vectors than that obtained here and therefore cannot index some of the superstructure peaks observed in the X-ray pattern, for example the peak at (2,~1/2,~0). 

\subsection{Angle-resolved photoemission spectroscopy}

\begin{figure*}
	\centering
	\includegraphics[width=\textwidth]{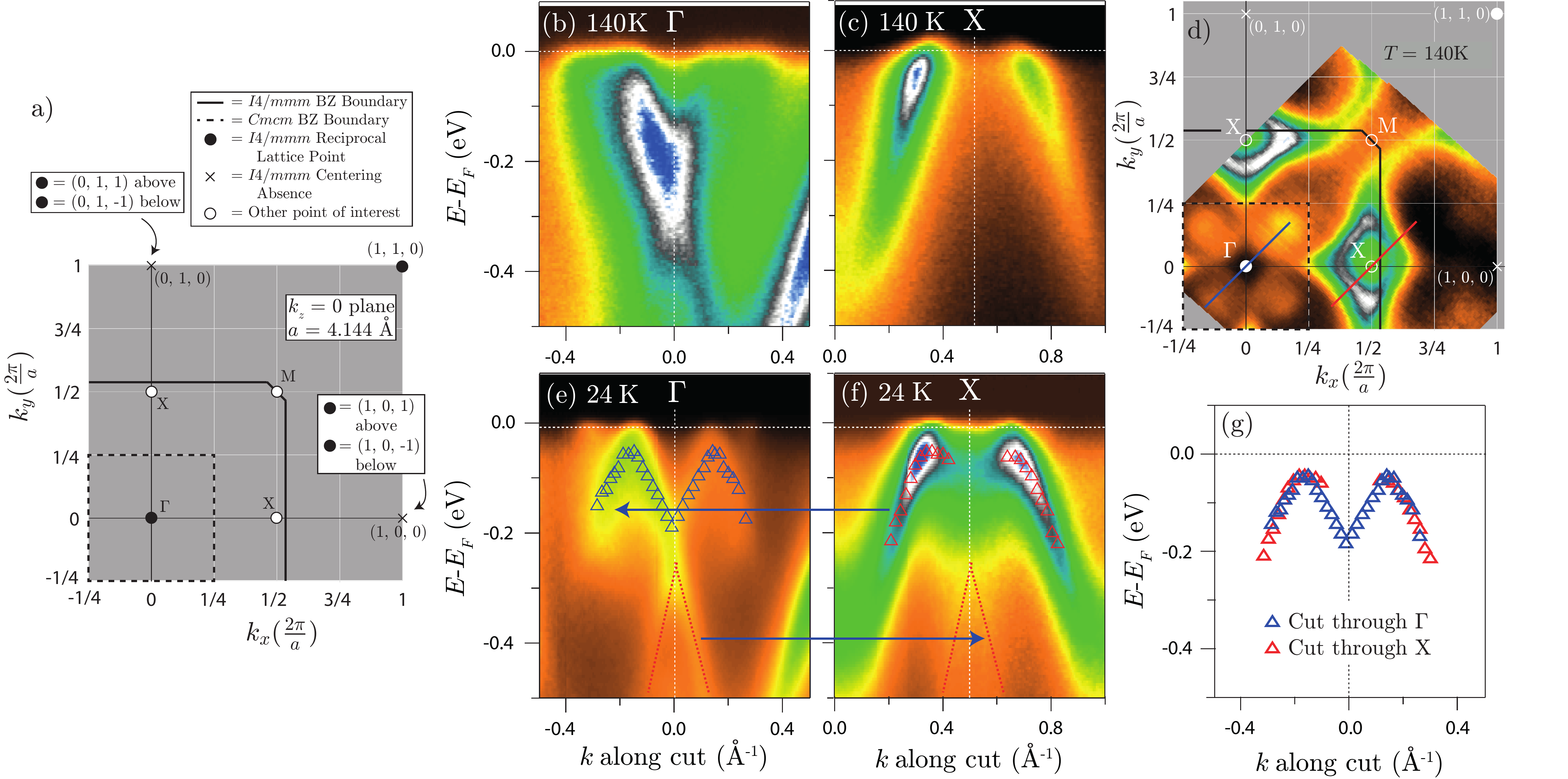}
	\caption{(a) $k_{z} = 0$ plane in the reciprocal space of Na$_{2}$Ti$_{2}$Sb$_{2}$O, with the folded ($Cmcm$) and unfolded ($I4/mmm$) Brillouin zone boundaries and the $\Gamma$, X and M points referred to in this figure as well as in the main text. (b, c, e and f) ARPES intensity plots of the band dispersions through the $\Gamma$ and X points at 140\,K and 24\,K, above and below $T_{\rm DW} \simeq 115$\,K, respectively. The momentum locations are indicated as blue (b, e) and red (c, f) lines in panel (d). Blue and red triangles mark the band dispersions at 24\,K determined from the energy distribution curves (EDCs). (d) Fermi surface intensity plot recorded at 140\,K. (g) Superimposition of the band dispersions in cuts through the $\Gamma$ [panel (e)] and X [panel (f)] points.}
	\label{ARPES_Fig}
\end{figure*}

ARPES measurements were performed on Na$_2$Ti$_2$Sb$_2$O in order to investigate the relationship between the lattice distortion, electronic structure and CDW. Figure~\ref{ARPES_Fig}(a) shows special points in the $k_{z} = 0$ plane in the reciprocal space of Na$_{2}$Ti$_{2}$Sb$_{2}$O. For the $I4/mmm$ parent structure, reciprocal lattice points (all symmetry-equivalent to $\Gamma = (0, 0, 0)$) are marked with filled circles and systematic absences with crosses. At the phase transition into the $Cmcm$ distorted structure determined by x-ray diffraction all crosses ($I4/mmm$ absences) become reciprocal lattice points, as well as the X = (1/2, 0, 0) and M = (1/2, 1/2, 0) points, which are marked with empty circles. According to our model, the X and $\Gamma$ points become equivalent in the $Cmcm$ structure below $T_{\rm DW}$, inducing a band folding from X to $\Gamma$.

Figures~\ref{ARPES_Fig}(b), \ref{ARPES_Fig}(c), \ref{ARPES_Fig}(e) and \ref{ARPES_Fig}(f) show measured band dispersions through the $\Gamma$ (b, e) and X (c, f) points above ($T = 140$\,K) and below ($T = 24$\,K) $T_{\rm DW}$, and Fig.~\ref{ARPES_Fig}(d) shows the observed Fermi surface (FS) in the undistorted state at 140\,K. The measured band structure at high temperature agrees with that reported in Ref.~\onlinecite{Tan2015} as well as with DFT calculations. In previous studies it has been established that there are three FS sheets, one around each of $\Gamma$, X and M, all of which are quasi-2D and have  approximately square cross-sections. The $\Gamma$ FS sheet is predicted to show a more pronounced 3D warping, although this has not yet been observed experimentally, perhaps due to limited $k_z$ resolution in the ARPES experiments.\cite{Tan2015} We observe all the $\Gamma$, X and M FS pockets in our dataset, although we see only two opposite sections of the squarish FS pocket at M, which we attribute to matrix element effects. Since these three FS pockets are similar in shape at the first order, it is plausible that a CDW with wavevectors ${\bf q}_1 = (1/2, 0, 0)$ and ${\bf q}_2 = (0, 1/2, 0)$ can develop due to nesting between them as suggested by Tan {\it et al.},\cite{Tan2015} depending on the precise 3D nature and warping of the FS sheets.

At 140\,K $ > T_{\rm DW}$, the band structures around $\Gamma$ and X are clearly different. While an electron band is found at $\Gamma$, a hole band is observed at the X point, both of which cross $E_{\rm F}$ directly (Fig.~\ref{ARPES_Fig}(b) and \ref{ARPES_Fig}(c)). Upon cooling down to 24\,K (Figs.~\ref{ARPES_Fig}(e), \ref{ARPES_Fig}(f) and \ref{ARPES_Fig}(g)) the $\Gamma$ and X bands fold onto each other and hybridize near $E_{\rm F}$, thus gapping the Fermi surface, as expected for a CDW gap formation. This band folding and the resulting back-bending can be seen in our data, particularly in Fig. \ref{ARPES_Fig}(g) where we extract the band positions relative to the nearby high-symmetry point \textit{via} Energy Distribution Curve (EDC) analysis. The folded bands coincide remarkably well, fully consistent with a CDW with the wavevectors ${\bf q}_1$ and ${\bf q}_2$. This provides evidence that at least some parts of the $\Gamma$ and X FS pockets are involved in the CDW gap formation. Whether the M point FS sheet is also involved, and whether the CDW gap on each of these three FS sheets is full or only partial as a function of $k_{z}$ remain open questions that may be answered \textit{via} more in-depth ARPES studies.

Our ARPES measurements cannot distinguish whether the CDW exhibits both or only one of the propagation vectors ${\bf q}_{1}$ and ${\bf q}_{2}$ due to the effects of domains in the single-${\bf q}$ case. However if CDWs with ${\bf q}_{1}$ and ${\bf q}_{2}$ do develop simultaneously (which is reasonable since ${\bf q}_{1}$ and ${\bf q}_2$ are equivalent in $I4/mmm$) then they are expected to couple to a two-${\bf q}$ lattice distortion with the same propagation vectors ${\bf q}_{1}$ and ${\bf q}_2$. We have detected precisely such a distortion in this work, providing a unified picture of a coupled CDW and lattice distortion appearing at $T_{\rm DW} = 115$\,K in Na$_{2}$Ti$_{2}$Sb$_{2}$O. ARPES data have not been published for the case of $Pn =$ As, but non-magnetic DFT calculations \cite{Yan2013} predict a very similar Fermi surface to $Pn =$ Sb with quasi-two-dimensional (2D) Fermi sheets around the $\Gamma$, X and M points. Nesting vectors of the form ${\bf q}_{1}=(1/2,0,l)$ and ${\bf q}_{2}=(0,1/2,l)$ for some $l$ leading to associated CDWs coupled to the lattice distortion are therefore plausible in this case too.

\section{Conclusion}
The identification of the DW transition in the titanium oxypnictides as a lattice distortion coupled to a CDW solves the puzzle of the nature of this transition and provides a complete determination of the low temperature ordered phase including the periodicity, symmetry and lattice distortion mode. Fermi surface nesting assists the CDW to form, and the opening of a gap on nested parts of the Fermi surface explains the increase in resistivity and drop in magnetic susceptibility observed at $T_{\rm DW}$. The observed lattice instability implies a strong electron-phonon coupling for at least this one mode, which points towards a conventional phonon-mediated mechanism for the superconducting phase found in doped BaTi$_{2}$Sb$_{2}$O. However, given that superconductivity appears upon suppression of the CDW transition, unconventional mechanisms involving charge fluctuations, similar to that proposed in Ref.~\onlinecite{Wang2015} for the copper oxide superconductors, cannot be ruled out. The results provide a new impetus for theoretical models of superconductivity in this system.

\section*{Acknowledgments}
This research was supported by the United Kingdom EPSRC (grant nos. EP/J012912/1 and EP/J017124/1), the Diamond Light Source, National Natural Science Foundation of China (nos. 11274367, 11474330, 11474340 and 11234014), the Chinese Academy of Sciences (nos.~XDB07000000 and XDB07010000), and the 973 project of the Ministry of Science and Technology of China (nos. 2011CB921701, 2011CBA00110 and 2015CB921301). RDJ acknowledges support from a Royal Society University Research Fellowship.

\bibliographystyle{apsrev4-1}
\bibliography{Na2Ti2Pn2O_Bibliography}

\begin{thebibliography}{54}%
\makeatletter
\providecommand \@ifxundefined [1]{%
 \@ifx{#1\undefined}
}%
\providecommand \@ifnum [1]{%
 \ifnum #1\expandafter \@firstoftwo
 \else \expandafter \@secondoftwo
 \fi
}%
\providecommand \@ifx [1]{%
 \ifx #1\expandafter \@firstoftwo
 \else \expandafter \@secondoftwo
 \fi
}%
\providecommand \natexlab [1]{#1}%
\providecommand \enquote  [1]{``#1''}%
\providecommand \bibnamefont  [1]{#1}%
\providecommand \bibfnamefont [1]{#1}%
\providecommand \citenamefont [1]{#1}%
\providecommand \href@noop [0]{\@secondoftwo}%
\providecommand \href [0]{\begingroup \@sanitize@url \@href}%
\providecommand \@href[1]{\@@startlink{#1}\@@href}%
\providecommand \@@href[1]{\endgroup#1\@@endlink}%
\providecommand \@sanitize@url [0]{\catcode `\\12\catcode `\$12\catcode
  `\&12\catcode `\#12\catcode `\^12\catcode `\_12\catcode `\%12\relax}%
\providecommand \@@startlink[1]{}%
\providecommand \@@endlink[0]{}%
\providecommand \url  [0]{\begingroup\@sanitize@url \@url }%
\providecommand \@url [1]{\endgroup\@href {#1}{\urlprefix }}%
\providecommand \urlprefix  [0]{URL }%
\providecommand \Eprint [0]{\href }%
\providecommand \doibase [0]{http://dx.doi.org/}%
\providecommand \selectlanguage [0]{\@gobble}%
\providecommand \bibinfo  [0]{\@secondoftwo}%
\providecommand \bibfield  [0]{\@secondoftwo}%
\providecommand \translation [1]{[#1]}%
\providecommand \BibitemOpen [0]{}%
\providecommand \bibitemStop [0]{}%
\providecommand \bibitemNoStop [0]{.\EOS\space}%
\providecommand \EOS [0]{\spacefactor3000\relax}%
\providecommand \BibitemShut  [1]{\csname bibitem#1\endcsname}%
\let\auto@bib@innerbib\@empty
\bibitem [{\citenamefont {Monthoux}\ \emph {et~al.}(2007)\citenamefont
  {Monthoux}, \citenamefont {Pines},\ and\ \citenamefont
  {Lonzarich}}]{Monthoux2007}%
  \BibitemOpen
  \bibfield  {author} {\bibinfo {author} {\bibfnamefont {P.}~\bibnamefont
  {Monthoux}}, \bibinfo {author} {\bibfnamefont {D.}~\bibnamefont {Pines}}, \
  and\ \bibinfo {author} {\bibfnamefont {G.~G.}\ \bibnamefont {Lonzarich}},\
  }\href {\doibase 10.1038/nature06480} {\bibfield  {journal} {\bibinfo
  {journal} {Nature}\ }\textbf {\bibinfo {volume} {450}},\ \bibinfo {pages}
  {1177} (\bibinfo {year} {2007})}\BibitemShut {NoStop}%
\bibitem [{\citenamefont {Mazin}\ \emph {et~al.}(2008)\citenamefont {Mazin},
  \citenamefont {Singh}, \citenamefont {Johannes},\ and\ \citenamefont
  {Du}}]{Mazin2008}%
  \BibitemOpen
  \bibfield  {author} {\bibinfo {author} {\bibfnamefont {I.~I.}\ \bibnamefont
  {Mazin}}, \bibinfo {author} {\bibfnamefont {D.~J.}\ \bibnamefont {Singh}},
  \bibinfo {author} {\bibfnamefont {M.~D.}\ \bibnamefont {Johannes}}, \ and\
  \bibinfo {author} {\bibfnamefont {M.~H.}\ \bibnamefont {Du}},\ }\href
  {\doibase 10.1103/PhysRevLett.101.057003} {\bibfield  {journal} {\bibinfo
  {journal} {Phys. Rev. Lett.}\ }\textbf {\bibinfo {volume} {101}},\ \bibinfo
  {pages} {057003} (\bibinfo {year} {2008})}\BibitemShut {NoStop}%
\bibitem [{\citenamefont {Scalapino}(2012)}]{Scalapino2012}%
  \BibitemOpen
  \bibfield  {author} {\bibinfo {author} {\bibfnamefont {D.~J.}\ \bibnamefont
  {Scalapino}},\ }\href {\doibase 10.1103/RevModPhys.84.1383} {\bibfield
  {journal} {\bibinfo  {journal} {Rev. Mod. Phys.}\ }\textbf {\bibinfo {volume}
  {84}},\ \bibinfo {pages} {1383} (\bibinfo {year} {2012})}\BibitemShut
  {NoStop}%
\bibitem [{\citenamefont {Chubukov}(2012)}]{Chubukov2012}%
  \BibitemOpen
  \bibfield  {author} {\bibinfo {author} {\bibfnamefont {A.}~\bibnamefont
  {Chubukov}},\ }\href {\doibase 10.1146/annurev-conmatphys-020911-125055}
  {\bibfield  {journal} {\bibinfo  {journal} {Annu. Rev. Condens. Matter
  Phys.}\ }\textbf {\bibinfo {volume} {3}},\ \bibinfo {pages} {57} (\bibinfo
  {year} {2012})}\BibitemShut {NoStop}%
\bibitem [{\citenamefont {Wilson}\ \emph {et~al.}(1975)\citenamefont {Wilson},
  \citenamefont {Di~Salvo},\ and\ \citenamefont {Mahajan}}]{Wilson1975}%
  \BibitemOpen
  \bibfield  {author} {\bibinfo {author} {\bibfnamefont {J.~A.}\ \bibnamefont
  {Wilson}}, \bibinfo {author} {\bibfnamefont {F.~J.}\ \bibnamefont
  {Di~Salvo}}, \ and\ \bibinfo {author} {\bibfnamefont {S.}~\bibnamefont
  {Mahajan}},\ }\href {\doibase 10.1080/00018737500101391} {\bibfield
  {journal} {\bibinfo  {journal} {Adv. Phys.}\ }\textbf {\bibinfo {volume}
  {24}},\ \bibinfo {pages} {117} (\bibinfo {year} {1975})}\BibitemShut
  {NoStop}%
\bibitem [{\citenamefont {Morosan}\ \emph {et~al.}(2006)\citenamefont
  {Morosan}, \citenamefont {Zandbergen}, \citenamefont {Dennis}, \citenamefont
  {Bos}, \citenamefont {Onose}, \citenamefont {Klimczuk}, \citenamefont
  {Ramirez}, \citenamefont {Ong},\ and\ \citenamefont {Cava}}]{Morosan2006}%
  \BibitemOpen
  \bibfield  {author} {\bibinfo {author} {\bibfnamefont {E.}~\bibnamefont
  {Morosan}}, \bibinfo {author} {\bibfnamefont {H.~W.}\ \bibnamefont
  {Zandbergen}}, \bibinfo {author} {\bibfnamefont {B.~S.}\ \bibnamefont
  {Dennis}}, \bibinfo {author} {\bibfnamefont {J.~W.~G.}\ \bibnamefont {Bos}},
  \bibinfo {author} {\bibfnamefont {Y.}~\bibnamefont {Onose}}, \bibinfo
  {author} {\bibfnamefont {T.}~\bibnamefont {Klimczuk}}, \bibinfo {author}
  {\bibfnamefont {A.~P.}\ \bibnamefont {Ramirez}}, \bibinfo {author}
  {\bibfnamefont {N.~P.}\ \bibnamefont {Ong}}, \ and\ \bibinfo {author}
  {\bibfnamefont {R.~J.}\ \bibnamefont {Cava}},\ }\href
  {http://dx.doi.org/10.1038/nphys360} {\bibfield  {journal} {\bibinfo
  {journal} {Nat. Phys.}\ }\textbf {\bibinfo {volume} {2}},\ \bibinfo {pages}
  {544} (\bibinfo {year} {2006})}\BibitemShut {NoStop}%
\bibitem [{\citenamefont {Yokoya}\ \emph {et~al.}(2005)\citenamefont {Yokoya},
  \citenamefont {Kiss}, \citenamefont {Chainani}, \citenamefont {Shin},\ and\
  \citenamefont {Yamaya}}]{Yokoya2005}%
  \BibitemOpen
  \bibfield  {author} {\bibinfo {author} {\bibfnamefont {T.}~\bibnamefont
  {Yokoya}}, \bibinfo {author} {\bibfnamefont {T.}~\bibnamefont {Kiss}},
  \bibinfo {author} {\bibfnamefont {A.}~\bibnamefont {Chainani}}, \bibinfo
  {author} {\bibfnamefont {S.}~\bibnamefont {Shin}}, \ and\ \bibinfo {author}
  {\bibfnamefont {K.}~\bibnamefont {Yamaya}},\ }\href {\doibase
  10.1103/PhysRevB.71.140504} {\bibfield  {journal} {\bibinfo  {journal} {Phys.
  Rev. B}\ }\textbf {\bibinfo {volume} {71}},\ \bibinfo {pages} {140504(R)}
  (\bibinfo {year} {2005})}\BibitemShut {NoStop}%
\bibitem [{\citenamefont {Mattheiss}\ and\ \citenamefont
  {Hamann}(1988)}]{Mattheiss1988}%
  \BibitemOpen
  \bibfield  {author} {\bibinfo {author} {\bibfnamefont {L.~F.}\ \bibnamefont
  {Mattheiss}}\ and\ \bibinfo {author} {\bibfnamefont {D.~R.}\ \bibnamefont
  {Hamann}},\ }\href {\doibase 10.1103/PhysRevLett.60.2681} {\bibfield
  {journal} {\bibinfo  {journal} {Phys. Rev. Lett.}\ }\textbf {\bibinfo
  {volume} {60}},\ \bibinfo {pages} {2681} (\bibinfo {year}
  {1988})}\BibitemShut {NoStop}%
\bibitem [{\citenamefont {Tranquada}\ \emph {et~al.}(1995)\citenamefont
  {Tranquada}, \citenamefont {Sternlieb}, \citenamefont {Axe}, \citenamefont
  {Nakamura},\ and\ \citenamefont {Uchida}}]{Tranquada1995}%
  \BibitemOpen
  \bibfield  {author} {\bibinfo {author} {\bibfnamefont {J.~M.}\ \bibnamefont
  {Tranquada}}, \bibinfo {author} {\bibfnamefont {B.~J.}\ \bibnamefont
  {Sternlieb}}, \bibinfo {author} {\bibfnamefont {J.~D.}\ \bibnamefont {Axe}},
  \bibinfo {author} {\bibfnamefont {Y.}~\bibnamefont {Nakamura}}, \ and\
  \bibinfo {author} {\bibfnamefont {S.}~\bibnamefont {Uchida}},\ }\href
  {http://dx.doi.org/10.1038/375561a0} {\bibfield  {journal} {\bibinfo
  {journal} {Nature}\ }\textbf {\bibinfo {volume} {375}},\ \bibinfo {pages}
  {561} (\bibinfo {year} {1995})}\BibitemShut {NoStop}%
\bibitem [{\citenamefont {Fujita}\ \emph {et~al.}(2004)\citenamefont {Fujita},
  \citenamefont {Goka}, \citenamefont {Yamada}, \citenamefont {Tranquada},\
  and\ \citenamefont {Regnault}}]{fujita2004}%
  \BibitemOpen
  \bibfield  {author} {\bibinfo {author} {\bibfnamefont {M.}~\bibnamefont
  {Fujita}}, \bibinfo {author} {\bibfnamefont {H.}~\bibnamefont {Goka}},
  \bibinfo {author} {\bibfnamefont {K.}~\bibnamefont {Yamada}}, \bibinfo
  {author} {\bibfnamefont {J.~M.}\ \bibnamefont {Tranquada}}, \ and\ \bibinfo
  {author} {\bibfnamefont {L.~P.}\ \bibnamefont {Regnault}},\ }\href {\doibase
  10.1103/PhysRevB.70.104517} {\bibfield  {journal} {\bibinfo  {journal} {Phys.
  Rev. B}\ }\textbf {\bibinfo {volume} {70}},\ \bibinfo {pages} {104517}
  (\bibinfo {year} {2004})}\BibitemShut {NoStop}%
\bibitem [{\citenamefont {Croft}\ \emph {et~al.}(2014)\citenamefont {Croft},
  \citenamefont {Lester}, \citenamefont {Senn}, \citenamefont {Bombardi},\ and\
  \citenamefont {Hayden}}]{Croft2014}%
  \BibitemOpen
  \bibfield  {author} {\bibinfo {author} {\bibfnamefont {T.~P.}\ \bibnamefont
  {Croft}}, \bibinfo {author} {\bibfnamefont {C.}~\bibnamefont {Lester}},
  \bibinfo {author} {\bibfnamefont {M.~S.}\ \bibnamefont {Senn}}, \bibinfo
  {author} {\bibfnamefont {A.}~\bibnamefont {Bombardi}}, \ and\ \bibinfo
  {author} {\bibfnamefont {S.~M.}\ \bibnamefont {Hayden}},\ }\href {\doibase
  10.1103/PhysRevB.89.224513} {\bibfield  {journal} {\bibinfo  {journal} {Phys.
  Rev. B}\ }\textbf {\bibinfo {volume} {89}},\ \bibinfo {pages} {224513}
  (\bibinfo {year} {2014})}\BibitemShut {NoStop}%
\bibitem [{\citenamefont {da~Silva~Neto}\ \emph {et~al.}(2014)\citenamefont
  {da~Silva~Neto}, \citenamefont {Aynajian}, \citenamefont {Frano},
  \citenamefont {Comin}, \citenamefont {Schierle}, \citenamefont {Weschke},
  \citenamefont {Gyenis}, \citenamefont {Wen}, \citenamefont {Schneeloch},
  \citenamefont {Xu}, \citenamefont {Ono}, \citenamefont {Gu}, \citenamefont
  {Le~Tacon},\ and\ \citenamefont {Yazdani}}]{daSilvaNeto2014}%
  \BibitemOpen
  \bibfield  {author} {\bibinfo {author} {\bibfnamefont {E.~H.}\ \bibnamefont
  {da~Silva~Neto}}, \bibinfo {author} {\bibfnamefont {P.}~\bibnamefont
  {Aynajian}}, \bibinfo {author} {\bibfnamefont {A.}~\bibnamefont {Frano}},
  \bibinfo {author} {\bibfnamefont {R.}~\bibnamefont {Comin}}, \bibinfo
  {author} {\bibfnamefont {E.}~\bibnamefont {Schierle}}, \bibinfo {author}
  {\bibfnamefont {E.}~\bibnamefont {Weschke}}, \bibinfo {author} {\bibfnamefont
  {A.}~\bibnamefont {Gyenis}}, \bibinfo {author} {\bibfnamefont
  {J.}~\bibnamefont {Wen}}, \bibinfo {author} {\bibfnamefont {J.}~\bibnamefont
  {Schneeloch}}, \bibinfo {author} {\bibfnamefont {Z.}~\bibnamefont {Xu}},
  \bibinfo {author} {\bibfnamefont {S.}~\bibnamefont {Ono}}, \bibinfo {author}
  {\bibfnamefont {G.}~\bibnamefont {Gu}}, \bibinfo {author} {\bibfnamefont
  {M.}~\bibnamefont {Le~Tacon}}, \ and\ \bibinfo {author} {\bibfnamefont
  {A.}~\bibnamefont {Yazdani}},\ }\href {\doibase 10.1126/science.1243479}
  {\bibfield  {journal} {\bibinfo  {journal} {Science}\ }\textbf {\bibinfo
  {volume} {343}},\ \bibinfo {pages} {393} (\bibinfo {year}
  {2014})}\BibitemShut {NoStop}%
\bibitem [{\citenamefont {Wu}\ \emph {et~al.}(2011)\citenamefont {Wu},
  \citenamefont {Mayaffre}, \citenamefont {Kr\"{a}mer}, \citenamefont
  {Horvati\'{c}}, \citenamefont {Berthier}, \citenamefont {Hardy},
  \citenamefont {Liang}, \citenamefont {Bonn},\ and\ \citenamefont
  {Julien}}]{Wu2011}%
  \BibitemOpen
  \bibfield  {author} {\bibinfo {author} {\bibfnamefont {T.}~\bibnamefont
  {Wu}}, \bibinfo {author} {\bibfnamefont {H.}~\bibnamefont {Mayaffre}},
  \bibinfo {author} {\bibfnamefont {S.}~\bibnamefont {Kr\"{a}mer}}, \bibinfo
  {author} {\bibfnamefont {M.}~\bibnamefont {Horvati\'{c}}}, \bibinfo {author}
  {\bibfnamefont {C.}~\bibnamefont {Berthier}}, \bibinfo {author}
  {\bibfnamefont {W.~N.}\ \bibnamefont {Hardy}}, \bibinfo {author}
  {\bibfnamefont {R.}~\bibnamefont {Liang}}, \bibinfo {author} {\bibfnamefont
  {D.~A.}\ \bibnamefont {Bonn}}, \ and\ \bibinfo {author} {\bibfnamefont
  {M.-H.}\ \bibnamefont {Julien}},\ }\href@noop {} {\bibfield  {journal}
  {\bibinfo  {journal} {Nature}\ }\textbf {\bibinfo {volume} {477}},\ \bibinfo
  {pages} {191} (\bibinfo {year} {2011})}\BibitemShut {NoStop}%
\bibitem [{\citenamefont {Ghiringhelli}\ \emph {et~al.}(2012)\citenamefont
  {Ghiringhelli}, \citenamefont {Le~Tacon}, \citenamefont {Minola},
  \citenamefont {Blanco-Canosa}, \citenamefont {Mazzoli}, \citenamefont
  {Brookes}, \citenamefont {De~Luca}, \citenamefont {Frano}, \citenamefont
  {Hawthorn}, \citenamefont {He}, \citenamefont {Loew}, \citenamefont
  {Moretti~Sala}, \citenamefont {Peets}, \citenamefont {Salluzo}, \citenamefont
  {Schierle}, \citenamefont {Sutarto}, \citenamefont {Sawatzky}, \citenamefont
  {Weschke}, \citenamefont {Keimer},\ and\ \citenamefont
  {Braicovich}}]{Ghiringhelli2012}%
  \BibitemOpen
  \bibfield  {author} {\bibinfo {author} {\bibfnamefont {G.}~\bibnamefont
  {Ghiringhelli}}, \bibinfo {author} {\bibfnamefont {M.}~\bibnamefont
  {Le~Tacon}}, \bibinfo {author} {\bibfnamefont {M.}~\bibnamefont {Minola}},
  \bibinfo {author} {\bibfnamefont {S.}~\bibnamefont {Blanco-Canosa}}, \bibinfo
  {author} {\bibfnamefont {C.}~\bibnamefont {Mazzoli}}, \bibinfo {author}
  {\bibfnamefont {N.~B.}\ \bibnamefont {Brookes}}, \bibinfo {author}
  {\bibfnamefont {G.~M.}\ \bibnamefont {De~Luca}}, \bibinfo {author}
  {\bibfnamefont {A.}~\bibnamefont {Frano}}, \bibinfo {author} {\bibfnamefont
  {D.~G.}\ \bibnamefont {Hawthorn}}, \bibinfo {author} {\bibfnamefont
  {F.}~\bibnamefont {He}}, \bibinfo {author} {\bibfnamefont {T.}~\bibnamefont
  {Loew}}, \bibinfo {author} {\bibfnamefont {M.}~\bibnamefont {Moretti~Sala}},
  \bibinfo {author} {\bibfnamefont {D.~C.}\ \bibnamefont {Peets}}, \bibinfo
  {author} {\bibfnamefont {M.}~\bibnamefont {Salluzo}}, \bibinfo {author}
  {\bibfnamefont {E.}~\bibnamefont {Schierle}}, \bibinfo {author}
  {\bibfnamefont {R.}~\bibnamefont {Sutarto}}, \bibinfo {author} {\bibfnamefont
  {G.~A.}\ \bibnamefont {Sawatzky}}, \bibinfo {author} {\bibfnamefont
  {E.}~\bibnamefont {Weschke}}, \bibinfo {author} {\bibfnamefont
  {B.}~\bibnamefont {Keimer}}, \ and\ \bibinfo {author} {\bibfnamefont
  {L.}~\bibnamefont {Braicovich}},\ }\href {\doibase 10.1126/science.1223532}
  {\bibfield  {journal} {\bibinfo  {journal} {Science}\ }\textbf {\bibinfo
  {volume} {337}},\ \bibinfo {pages} {821} (\bibinfo {year}
  {2012})}\BibitemShut {NoStop}%
\bibitem [{\citenamefont {Chang}\ \emph {et~al.}(2012)\citenamefont {Chang},
  \citenamefont {Blackburn}, \citenamefont {Holmes}, \citenamefont
  {Christensen}, \citenamefont {Larsen}, \citenamefont {Mesot}, \citenamefont
  {Liang}, \citenamefont {Bonn}, \citenamefont {Hardy}, \citenamefont
  {Watenphul}, \citenamefont {Zimmermann}, \citenamefont {Forgan},\ and\
  \citenamefont {Hayden}}]{Chang2012}%
  \BibitemOpen
  \bibfield  {author} {\bibinfo {author} {\bibfnamefont {J.}~\bibnamefont
  {Chang}}, \bibinfo {author} {\bibfnamefont {E.}~\bibnamefont {Blackburn}},
  \bibinfo {author} {\bibfnamefont {A.~T.}\ \bibnamefont {Holmes}}, \bibinfo
  {author} {\bibfnamefont {N.~B.}\ \bibnamefont {Christensen}}, \bibinfo
  {author} {\bibfnamefont {J.}~\bibnamefont {Larsen}}, \bibinfo {author}
  {\bibfnamefont {J.}~\bibnamefont {Mesot}}, \bibinfo {author} {\bibfnamefont
  {R.}~\bibnamefont {Liang}}, \bibinfo {author} {\bibfnamefont {D.~A.}\
  \bibnamefont {Bonn}}, \bibinfo {author} {\bibfnamefont {W.~N.}\ \bibnamefont
  {Hardy}}, \bibinfo {author} {\bibfnamefont {A.}~\bibnamefont {Watenphul}},
  \bibinfo {author} {\bibfnamefont {M.~v.}\ \bibnamefont {Zimmermann}},
  \bibinfo {author} {\bibfnamefont {E.~M.}\ \bibnamefont {Forgan}}, \ and\
  \bibinfo {author} {\bibfnamefont {S.~M.}\ \bibnamefont {Hayden}},\ }\href
  {http://dx.doi.org/10.1038/nphys2456} {\bibfield  {journal} {\bibinfo
  {journal} {Nat. Phys.}\ }\textbf {\bibinfo {volume} {8}},\ \bibinfo {pages}
  {871} (\bibinfo {year} {2012})}\BibitemShut {NoStop}%
\bibitem [{\citenamefont {Wang}\ and\ \citenamefont
  {Chubukov}(2015)}]{Wang2015}%
  \BibitemOpen
  \bibfield  {author} {\bibinfo {author} {\bibfnamefont {Y.}~\bibnamefont
  {Wang}}\ and\ \bibinfo {author} {\bibfnamefont {A.~V.}\ \bibnamefont
  {Chubukov}},\ }\href {\doibase 10.1103/PhysRevB.92.125108} {\bibfield
  {journal} {\bibinfo  {journal} {Phys. Rev. B}\ }\textbf {\bibinfo {volume}
  {92}},\ \bibinfo {pages} {125108} (\bibinfo {year} {2015})}\BibitemShut
  {NoStop}%
\bibitem [{\citenamefont {Lorenz}\ \emph {et~al.}(2014)\citenamefont {Lorenz},
  \citenamefont {Guloy},\ and\ \citenamefont {Chu}}]{Lorenz2014}%
  \BibitemOpen
  \bibfield  {author} {\bibinfo {author} {\bibfnamefont {B.}~\bibnamefont
  {Lorenz}}, \bibinfo {author} {\bibfnamefont {A.~M.}\ \bibnamefont {Guloy}}, \
  and\ \bibinfo {author} {\bibfnamefont {P.~C.~W.}\ \bibnamefont {Chu}},\
  }\href {\doibase 10.1142/S0217979214300114} {\bibfield  {journal} {\bibinfo
  {journal} {Int. J. Mod. Phys. B}\ }\textbf {\bibinfo {volume} {28}},\
  \bibinfo {pages} {1430011} (\bibinfo {year} {2014})}\BibitemShut {NoStop}%
\bibitem [{\citenamefont {Adam}\ and\ \citenamefont
  {Schuster}(1990)}]{Adam1990}%
  \BibitemOpen
  \bibfield  {author} {\bibinfo {author} {\bibfnamefont {A.}~\bibnamefont
  {Adam}}\ and\ \bibinfo {author} {\bibfnamefont {H.-U.}\ \bibnamefont
  {Schuster}},\ }\href {\doibase 10.1002/zaac.19905840115} {\bibfield
  {journal} {\bibinfo  {journal} {Z. Anorg. Allg. Chem.}\ }\textbf {\bibinfo
  {volume} {584}},\ \bibinfo {pages} {150} (\bibinfo {year}
  {1990})}\BibitemShut {NoStop}%
\bibitem [{\citenamefont {Axtell}\ \emph {et~al.}(1997)\citenamefont {Axtell},
  \citenamefont {Ozawa}, \citenamefont {Kauzlarich},\ and\ \citenamefont
  {Singh}}]{Axtell1997}%
  \BibitemOpen
  \bibfield  {author} {\bibinfo {author} {\bibfnamefont {E.~A.}\ \bibnamefont
  {Axtell}, \bibfnamefont {III}}, \bibinfo {author} {\bibfnamefont
  {T.}~\bibnamefont {Ozawa}}, \bibinfo {author} {\bibfnamefont {S.~M.}\
  \bibnamefont {Kauzlarich}}, \ and\ \bibinfo {author} {\bibfnamefont
  {R.~R.~P.}\ \bibnamefont {Singh}},\ }\href {\doibase
  http://dx.doi.org/10.1006/jssc.1997.7715} {\bibfield  {journal} {\bibinfo
  {journal} {J. Solid State Chem.}\ }\textbf {\bibinfo {volume} {134}},\
  \bibinfo {pages} {423 } (\bibinfo {year} {1997})}\BibitemShut {NoStop}%
\bibitem [{\citenamefont {Ozawa}\ \emph {et~al.}(2000)\citenamefont {Ozawa},
  \citenamefont {Pantoja}, \citenamefont {Axtell}, \citenamefont {Kauzlarich},
  \citenamefont {Greedan}, \citenamefont {Bieringer},\ and\ \citenamefont
  {Richardson~Jr.}}]{Ozawa2000}%
  \BibitemOpen
  \bibfield  {author} {\bibinfo {author} {\bibfnamefont {T.~C.}\ \bibnamefont
  {Ozawa}}, \bibinfo {author} {\bibfnamefont {R.}~\bibnamefont {Pantoja}},
  \bibinfo {author} {\bibfnamefont {E.~A.}\ \bibnamefont {Axtell},
  \bibfnamefont {III}}, \bibinfo {author} {\bibfnamefont {S.~M.}\ \bibnamefont
  {Kauzlarich}}, \bibinfo {author} {\bibfnamefont {J.~E.}\ \bibnamefont
  {Greedan}}, \bibinfo {author} {\bibfnamefont {M.}~\bibnamefont {Bieringer}},
  \ and\ \bibinfo {author} {\bibfnamefont {J.~W.}\ \bibnamefont
  {Richardson~Jr.}},\ }\href {\doibase
  http://dx.doi.org/10.1006/jssc.2000.8758} {\bibfield  {journal} {\bibinfo
  {journal} {J. Solid State Chem.}\ }\textbf {\bibinfo {volume} {153}},\
  \bibinfo {pages} {275} (\bibinfo {year} {2000})}\BibitemShut {NoStop}%
\bibitem [{\citenamefont {Ozawa}\ \emph {et~al.}(2001)\citenamefont {Ozawa},
  \citenamefont {Kauzlarich}, \citenamefont {Bieringer},\ and\ \citenamefont
  {Greedan}}]{Ozawa2001}%
  \BibitemOpen
  \bibfield  {author} {\bibinfo {author} {\bibfnamefont {T.~C.}\ \bibnamefont
  {Ozawa}}, \bibinfo {author} {\bibfnamefont {S.~M.}\ \bibnamefont
  {Kauzlarich}}, \bibinfo {author} {\bibfnamefont {M.}~\bibnamefont
  {Bieringer}}, \ and\ \bibinfo {author} {\bibfnamefont {J.~E.}\ \bibnamefont
  {Greedan}},\ }\href {\doibase 10.1021/cm010009f} {\bibfield  {journal}
  {\bibinfo  {journal} {Chem. Mater.}\ }\textbf {\bibinfo {volume} {13}},\
  \bibinfo {pages} {1804} (\bibinfo {year} {2001})}\BibitemShut {NoStop}%
\bibitem [{\citenamefont {Liu}\ \emph {et~al.}(2009)\citenamefont {Liu},
  \citenamefont {Tan}, \citenamefont {Song}, \citenamefont {Li}, \citenamefont
  {Yan}, \citenamefont {Ying}, \citenamefont {Xie}, \citenamefont {Wang},\ and\
  \citenamefont {Chen}}]{Liu2009}%
  \BibitemOpen
  \bibfield  {author} {\bibinfo {author} {\bibfnamefont {R.~H.}\ \bibnamefont
  {Liu}}, \bibinfo {author} {\bibfnamefont {D.}~\bibnamefont {Tan}}, \bibinfo
  {author} {\bibfnamefont {Y.~A.}\ \bibnamefont {Song}}, \bibinfo {author}
  {\bibfnamefont {Q.~J.}\ \bibnamefont {Li}}, \bibinfo {author} {\bibfnamefont
  {Y.~J.}\ \bibnamefont {Yan}}, \bibinfo {author} {\bibfnamefont {J.~J.}\
  \bibnamefont {Ying}}, \bibinfo {author} {\bibfnamefont {Y.~L.}\ \bibnamefont
  {Xie}}, \bibinfo {author} {\bibfnamefont {X.~F.}\ \bibnamefont {Wang}}, \
  and\ \bibinfo {author} {\bibfnamefont {X.~H.}\ \bibnamefont {Chen}},\ }\href
  {\doibase 10.1103/PhysRevB.80.144516} {\bibfield  {journal} {\bibinfo
  {journal} {Phys. Rev. B}\ }\textbf {\bibinfo {volume} {80}},\ \bibinfo
  {pages} {144516} (\bibinfo {year} {2009})}\BibitemShut {NoStop}%
\bibitem [{\citenamefont {Liu}\ \emph {et~al.}(2010)\citenamefont {Liu},
  \citenamefont {Song}, \citenamefont {Li}, \citenamefont {Ying}, \citenamefont
  {Yan}, \citenamefont {He},\ and\ \citenamefont {Chen}}]{Liu2010}%
  \BibitemOpen
  \bibfield  {author} {\bibinfo {author} {\bibfnamefont {R.~H.}\ \bibnamefont
  {Liu}}, \bibinfo {author} {\bibfnamefont {Y.~A.}\ \bibnamefont {Song}},
  \bibinfo {author} {\bibfnamefont {Q.~J.}\ \bibnamefont {Li}}, \bibinfo
  {author} {\bibfnamefont {J.~J.}\ \bibnamefont {Ying}}, \bibinfo {author}
  {\bibfnamefont {Y.~J.}\ \bibnamefont {Yan}}, \bibinfo {author} {\bibfnamefont
  {Y.}~\bibnamefont {He}}, \ and\ \bibinfo {author} {\bibfnamefont {X.~H.}\
  \bibnamefont {Chen}},\ }\href {\doibase 10.1021/cm9027258} {\bibfield
  {journal} {\bibinfo  {journal} {Chem. Mater.}\ }\textbf {\bibinfo {volume}
  {22}},\ \bibinfo {pages} {1503} (\bibinfo {year} {2010})}\BibitemShut
  {NoStop}%
\bibitem [{\citenamefont {Wang}\ \emph {et~al.}(2010)\citenamefont {Wang},
  \citenamefont {Yan}, \citenamefont {Ying}, \citenamefont {Li}, \citenamefont
  {Zhang}, \citenamefont {Xu},\ and\ \citenamefont {Chen}}]{Wang2010}%
  \BibitemOpen
  \bibfield  {author} {\bibinfo {author} {\bibfnamefont {X.~F.}\ \bibnamefont
  {Wang}}, \bibinfo {author} {\bibfnamefont {Y.~J.}\ \bibnamefont {Yan}},
  \bibinfo {author} {\bibfnamefont {J.~J.}\ \bibnamefont {Ying}}, \bibinfo
  {author} {\bibfnamefont {Q.~J.}\ \bibnamefont {Li}}, \bibinfo {author}
  {\bibfnamefont {M.}~\bibnamefont {Zhang}}, \bibinfo {author} {\bibfnamefont
  {N.}~\bibnamefont {Xu}}, \ and\ \bibinfo {author} {\bibfnamefont {X.~H.}\
  \bibnamefont {Chen}},\ }\href@noop {} {\bibfield  {journal} {\bibinfo
  {journal} {J. Phys.: Condens. Matter}\ }\textbf {\bibinfo {volume} {22}},\
  \bibinfo {pages} {075702} (\bibinfo {year} {2010})}\BibitemShut {NoStop}%
\bibitem [{\citenamefont {Yajima}\ \emph {et~al.}(2012)\citenamefont {Yajima},
  \citenamefont {Nakano}, \citenamefont {Takeiri}, \citenamefont {Ono},
  \citenamefont {Hosokoshi}, \citenamefont {Matsushita}, \citenamefont
  {Hester}, \citenamefont {Kobayashi},\ and\ \citenamefont
  {Kageyama}}]{Yajima2012}%
  \BibitemOpen
  \bibfield  {author} {\bibinfo {author} {\bibfnamefont {T.}~\bibnamefont
  {Yajima}}, \bibinfo {author} {\bibfnamefont {K.}~\bibnamefont {Nakano}},
  \bibinfo {author} {\bibfnamefont {F.}~\bibnamefont {Takeiri}}, \bibinfo
  {author} {\bibfnamefont {T.}~\bibnamefont {Ono}}, \bibinfo {author}
  {\bibfnamefont {Y.}~\bibnamefont {Hosokoshi}}, \bibinfo {author}
  {\bibfnamefont {Y.}~\bibnamefont {Matsushita}}, \bibinfo {author}
  {\bibfnamefont {J.}~\bibnamefont {Hester}}, \bibinfo {author} {\bibfnamefont
  {Y.}~\bibnamefont {Kobayashi}}, \ and\ \bibinfo {author} {\bibfnamefont
  {H.}~\bibnamefont {Kageyama}},\ }\href {\doibase 10.1143/JPSJ.81.103706}
  {\bibfield  {journal} {\bibinfo  {journal} {J. Phys. Soc. of Jpn.}\ }\textbf
  {\bibinfo {volume} {81}},\ \bibinfo {pages} {103706} (\bibinfo {year}
  {2012})}\BibitemShut {NoStop}%
\bibitem [{\citenamefont {Doan}\ \emph {et~al.}(2012)\citenamefont {Doan},
  \citenamefont {Gooch}, \citenamefont {Tang}, \citenamefont {Lorenz},
  \citenamefont {M\"{o}ller}, \citenamefont {Tapp}, \citenamefont {Chu},\ and\
  \citenamefont {Guloy}}]{Doan2012}%
  \BibitemOpen
  \bibfield  {author} {\bibinfo {author} {\bibfnamefont {P.}~\bibnamefont
  {Doan}}, \bibinfo {author} {\bibfnamefont {M.}~\bibnamefont {Gooch}},
  \bibinfo {author} {\bibfnamefont {Z.}~\bibnamefont {Tang}}, \bibinfo {author}
  {\bibfnamefont {B.}~\bibnamefont {Lorenz}}, \bibinfo {author} {\bibfnamefont
  {A.}~\bibnamefont {M\"{o}ller}}, \bibinfo {author} {\bibfnamefont
  {J.}~\bibnamefont {Tapp}}, \bibinfo {author} {\bibfnamefont {P.~C.~W.}\
  \bibnamefont {Chu}}, \ and\ \bibinfo {author} {\bibfnamefont {A.~M.}\
  \bibnamefont {Guloy}},\ }\href {\doibase 10.1021/ja3078889} {\bibfield
  {journal} {\bibinfo  {journal} {J. Am. Chem. Soc.}\ }\textbf {\bibinfo
  {volume} {134}},\ \bibinfo {pages} {16520} (\bibinfo {year}
  {2012})}\BibitemShut {NoStop}%
\bibitem [{\citenamefont {Yajima}\ \emph {et~al.}(2013)\citenamefont {Yajima},
  \citenamefont {Nakano}, \citenamefont {Takeiri}, \citenamefont {Nozaki},
  \citenamefont {Kobayashi},\ and\ \citenamefont {Kageyama}}]{Yajima2013}%
  \BibitemOpen
  \bibfield  {author} {\bibinfo {author} {\bibfnamefont {T.}~\bibnamefont
  {Yajima}}, \bibinfo {author} {\bibfnamefont {K.}~\bibnamefont {Nakano}},
  \bibinfo {author} {\bibfnamefont {F.}~\bibnamefont {Takeiri}}, \bibinfo
  {author} {\bibfnamefont {Y.}~\bibnamefont {Nozaki}}, \bibinfo {author}
  {\bibfnamefont {Y.}~\bibnamefont {Kobayashi}}, \ and\ \bibinfo {author}
  {\bibfnamefont {H.}~\bibnamefont {Kageyama}},\ }\href {\doibase
  10.7566/JPSJ.82.033705} {\bibfield  {journal} {\bibinfo  {journal} {J. Phys.
  Soc. Jpn.}\ }\textbf {\bibinfo {volume} {82}},\ \bibinfo {pages} {033705}
  (\bibinfo {year} {2013})}\BibitemShut {NoStop}%
\bibitem [{\citenamefont {Zhai}\ \emph {et~al.}(2013)\citenamefont {Zhai},
  \citenamefont {Jiao}, \citenamefont {Sun}, \citenamefont {Bao}, \citenamefont
  {Jiang}, \citenamefont {Yang}, \citenamefont {Tang}, \citenamefont {Tao},
  \citenamefont {Xu}, \citenamefont {Li}, \citenamefont {Cao}, \citenamefont
  {Dai}, \citenamefont {Xu},\ and\ \citenamefont {Cao}}]{Zhai2013}%
  \BibitemOpen
  \bibfield  {author} {\bibinfo {author} {\bibfnamefont {H.~F.}\ \bibnamefont
  {Zhai}}, \bibinfo {author} {\bibfnamefont {W.~H.}\ \bibnamefont {Jiao}},
  \bibinfo {author} {\bibfnamefont {Y.~L.}\ \bibnamefont {Sun}}, \bibinfo
  {author} {\bibfnamefont {J.~K.}\ \bibnamefont {Bao}}, \bibinfo {author}
  {\bibfnamefont {H.}~\bibnamefont {Jiang}}, \bibinfo {author} {\bibfnamefont
  {X.~J.}\ \bibnamefont {Yang}}, \bibinfo {author} {\bibfnamefont {Z.~T.}\
  \bibnamefont {Tang}}, \bibinfo {author} {\bibfnamefont {Q.}~\bibnamefont
  {Tao}}, \bibinfo {author} {\bibfnamefont {X.~F.}\ \bibnamefont {Xu}},
  \bibinfo {author} {\bibfnamefont {Y.~K.}\ \bibnamefont {Li}}, \bibinfo
  {author} {\bibfnamefont {C.}~\bibnamefont {Cao}}, \bibinfo {author}
  {\bibfnamefont {J.~H.}\ \bibnamefont {Dai}}, \bibinfo {author} {\bibfnamefont
  {Z.~A.}\ \bibnamefont {Xu}}, \ and\ \bibinfo {author} {\bibfnamefont {G.~H.}\
  \bibnamefont {Cao}},\ }\href {\doibase 10.1103/PhysRevB.87.100502} {\bibfield
   {journal} {\bibinfo  {journal} {Phys. Rev. B}\ }\textbf {\bibinfo {volume}
  {87}},\ \bibinfo {pages} {100502(R)} (\bibinfo {year} {2013})}\BibitemShut
  {NoStop}%
\bibitem [{\citenamefont {Nakano}\ \emph {et~al.}(2013)\citenamefont {Nakano},
  \citenamefont {Yajima}, \citenamefont {Takeiri}, \citenamefont {Green},
  \citenamefont {Hester}, \citenamefont {Kobayashi},\ and\ \citenamefont
  {Kageyama}}]{Nakano2013}%
  \BibitemOpen
  \bibfield  {author} {\bibinfo {author} {\bibfnamefont {K.}~\bibnamefont
  {Nakano}}, \bibinfo {author} {\bibfnamefont {T.}~\bibnamefont {Yajima}},
  \bibinfo {author} {\bibfnamefont {F.}~\bibnamefont {Takeiri}}, \bibinfo
  {author} {\bibfnamefont {M.~A.}\ \bibnamefont {Green}}, \bibinfo {author}
  {\bibfnamefont {J.}~\bibnamefont {Hester}}, \bibinfo {author} {\bibfnamefont
  {Y.}~\bibnamefont {Kobayashi}}, \ and\ \bibinfo {author} {\bibfnamefont
  {H.}~\bibnamefont {Kageyama}},\ }\href {\doibase 10.7566/JPSJ.82.074707}
  {\bibfield  {journal} {\bibinfo  {journal} {J. Phys. Soc. Jpn.}\ }\textbf
  {\bibinfo {volume} {82}},\ \bibinfo {pages} {074707} (\bibinfo {year}
  {2013})}\BibitemShut {NoStop}%
\bibitem [{\citenamefont {Pachmayr}\ and\ \citenamefont
  {Johrendt}(2014)}]{Pachmayr2014}%
  \BibitemOpen
  \bibfield  {author} {\bibinfo {author} {\bibfnamefont {U.}~\bibnamefont
  {Pachmayr}}\ and\ \bibinfo {author} {\bibfnamefont {D.}~\bibnamefont
  {Johrendt}},\ }\href {\doibase 10.1016/j.solidstatesciences.2013.12.005}
  {\bibfield  {journal} {\bibinfo  {journal} {Solid State Sci.}\ }\textbf
  {\bibinfo {volume} {28}},\ \bibinfo {pages} {31} (\bibinfo {year}
  {2014})}\BibitemShut {NoStop}%
\bibitem [{\citenamefont {von Rohr}\ \emph {et~al.}(2014)\citenamefont {von
  Rohr}, \citenamefont {Nesper},\ and\ \citenamefont
  {Schilling}}]{vonRohr2014}%
  \BibitemOpen
  \bibfield  {author} {\bibinfo {author} {\bibfnamefont {F.}~\bibnamefont {von
  Rohr}}, \bibinfo {author} {\bibfnamefont {R.}~\bibnamefont {Nesper}}, \ and\
  \bibinfo {author} {\bibfnamefont {A.}~\bibnamefont {Schilling}},\ }\href
  {\doibase 10.1103/PhysRevB.89.094505} {\bibfield  {journal} {\bibinfo
  {journal} {Phys. Rev. B}\ }\textbf {\bibinfo {volume} {89}},\ \bibinfo
  {pages} {094505} (\bibinfo {year} {2014})}\BibitemShut {NoStop}%
\bibitem [{\citenamefont {Pickett}(1998)}]{Pickett1998}%
  \BibitemOpen
  \bibfield  {author} {\bibinfo {author} {\bibfnamefont {W.~E.}\ \bibnamefont
  {Pickett}},\ }\href {\doibase 10.1103/PhysRevB.58.4335} {\bibfield  {journal}
  {\bibinfo  {journal} {Phys. Rev. B}\ }\textbf {\bibinfo {volume} {58}},\
  \bibinfo {pages} {4335} (\bibinfo {year} {1998})}\BibitemShut {NoStop}%
\bibitem [{\citenamefont {de~Biani}\ \emph {et~al.}(1998)\citenamefont
  {de~Biani}, \citenamefont {Alemany},\ and\ \citenamefont
  {Canadell}}]{deBiani1998}%
  \BibitemOpen
  \bibfield  {author} {\bibinfo {author} {\bibfnamefont {F.~F.}\ \bibnamefont
  {de~Biani}}, \bibinfo {author} {\bibfnamefont {P.}~\bibnamefont {Alemany}}, \
  and\ \bibinfo {author} {\bibfnamefont {E.}~\bibnamefont {Canadell}},\
  }\href@noop {} {\bibfield  {journal} {\bibinfo  {journal} {Inorg. Chem.}\
  }\textbf {\bibinfo {volume} {37}},\ \bibinfo {pages} {5807} (\bibinfo {year}
  {1998})}\BibitemShut {NoStop}%
\bibitem [{\citenamefont {Singh}(2012)}]{Singh2012}%
  \BibitemOpen
  \bibfield  {author} {\bibinfo {author} {\bibfnamefont {D.~J.}\ \bibnamefont
  {Singh}},\ }\href {http://stacks.iop.org/1367-2630/14/i=12/a=123003}
  {\bibfield  {journal} {\bibinfo  {journal} {New J. Phys.}\ }\textbf {\bibinfo
  {volume} {14}},\ \bibinfo {pages} {123003} (\bibinfo {year}
  {2012})}\BibitemShut {NoStop}%
\bibitem [{\citenamefont {Wang}\ \emph {et~al.}(2013)\citenamefont {Wang},
  \citenamefont {Zhang}, \citenamefont {Zhang},\ and\ \citenamefont
  {Liu}}]{Wang2013}%
  \BibitemOpen
  \bibfield  {author} {\bibinfo {author} {\bibfnamefont {G.}~\bibnamefont
  {Wang}}, \bibinfo {author} {\bibfnamefont {H.}~\bibnamefont {Zhang}},
  \bibinfo {author} {\bibfnamefont {L.}~\bibnamefont {Zhang}}, \ and\ \bibinfo
  {author} {\bibfnamefont {C.}~\bibnamefont {Liu}},\ }\href@noop {} {\bibfield
  {journal} {\bibinfo  {journal} {J. Appl. Phys.}\ }\textbf {\bibinfo {volume}
  {113}},\ \bibinfo {pages} {243904} (\bibinfo {year} {2013})}\BibitemShut
  {NoStop}%
\bibitem [{\citenamefont {Yan}\ and\ \citenamefont {Lu}(2013)}]{Yan2013}%
  \BibitemOpen
  \bibfield  {author} {\bibinfo {author} {\bibfnamefont {X.~W.}\ \bibnamefont
  {Yan}}\ and\ \bibinfo {author} {\bibfnamefont {Z.~Y.}\ \bibnamefont {Lu}},\
  }\href {http://stacks.iop.org/0953-8984/25/i=36/a=365501} {\bibfield
  {journal} {\bibinfo  {journal} {J. Phys.: Condens. Matter}\ }\textbf
  {\bibinfo {volume} {25}},\ \bibinfo {pages} {365501} (\bibinfo {year}
  {2013})}\BibitemShut {NoStop}%
\bibitem [{\citenamefont {Suetin}\ and\ \citenamefont
  {Ivanovskii}(2013)}]{Suetin2013}%
  \BibitemOpen
  \bibfield  {author} {\bibinfo {author} {\bibfnamefont {D.~V.}\ \bibnamefont
  {Suetin}}\ and\ \bibinfo {author} {\bibfnamefont {A.~L.}\ \bibnamefont
  {Ivanovskii}},\ }\href@noop {} {\bibfield  {journal} {\bibinfo  {journal} {J.
  Alloys Comp.}\ }\textbf {\bibinfo {volume} {564}},\ \bibinfo {pages} {117}
  (\bibinfo {year} {2013})}\BibitemShut {NoStop}%
\bibitem [{\citenamefont {Yu}\ \emph {et~al.}(2014)\citenamefont {Yu},
  \citenamefont {Liu}, \citenamefont {Quan}, \citenamefont {Jia},\ and\
  \citenamefont {Lin}}]{Yu2014}%
  \BibitemOpen
  \bibfield  {author} {\bibinfo {author} {\bibfnamefont {X.~L.}\ \bibnamefont
  {Yu}}, \bibinfo {author} {\bibfnamefont {D.~Y.}\ \bibnamefont {Liu}},
  \bibinfo {author} {\bibfnamefont {Y.~M.}\ \bibnamefont {Quan}}, \bibinfo
  {author} {\bibfnamefont {T.}~\bibnamefont {Jia}}, \ and\ \bibinfo {author}
  {\bibfnamefont {H.~Q.}\ \bibnamefont {Lin}},\ }\href {\doibase
  10.1063/1.4866701} {\bibfield  {journal} {\bibinfo  {journal} {J. Appl.
  Phys.}\ }\textbf {\bibinfo {volume} {115}},\ \bibinfo {pages} {17A924}
  (\bibinfo {year} {2014})}\BibitemShut {NoStop}%
\bibitem [{\citenamefont {Subedi}(2013)}]{Subedi2013}%
  \BibitemOpen
  \bibfield  {author} {\bibinfo {author} {\bibfnamefont {A.}~\bibnamefont
  {Subedi}},\ }\href {\doibase 10.1103/PhysRevB.87.054506} {\bibfield
  {journal} {\bibinfo  {journal} {Phys. Rev. B}\ }\textbf {\bibinfo {volume}
  {87}},\ \bibinfo {pages} {054506} (\bibinfo {year} {2013})}\BibitemShut
  {NoStop}%
\bibitem [{\citenamefont {Nakano}\ \emph {et~al.}(2016)\citenamefont {Nakano},
  \citenamefont {Hongo},\ and\ \citenamefont {Maezono}}]{Nakano2016}%
  \BibitemOpen
  \bibfield  {author} {\bibinfo {author} {\bibfnamefont {K.}~\bibnamefont
  {Nakano}}, \bibinfo {author} {\bibfnamefont {K.}~\bibnamefont {Hongo}}, \
  and\ \bibinfo {author} {\bibfnamefont {R.}~\bibnamefont {Maezono}},\ }\href
  {http://dx.doi.org/10.1038/srep29661} {\bibfield  {journal} {\bibinfo
  {journal} {Scientific Reports}\ }\textbf {\bibinfo {volume} {6}},\ \bibinfo
  {pages} {29661} (\bibinfo {year} {2016})}\BibitemShut {NoStop}%
\bibitem [{\citenamefont {Tan}\ \emph {et~al.}(2015)\citenamefont {Tan},
  \citenamefont {Jiang}, \citenamefont {Ye}, \citenamefont {Niu}, \citenamefont
  {Song}, \citenamefont {Zhang}, \citenamefont {Dai}, \citenamefont {Xie},
  \citenamefont {Lai},\ and\ \citenamefont {Feng}}]{Tan2015}%
  \BibitemOpen
  \bibfield  {author} {\bibinfo {author} {\bibfnamefont {S.~Y.}\ \bibnamefont
  {Tan}}, \bibinfo {author} {\bibfnamefont {J.}~\bibnamefont {Jiang}}, \bibinfo
  {author} {\bibfnamefont {Z.~R.}\ \bibnamefont {Ye}}, \bibinfo {author}
  {\bibfnamefont {X.~H.}\ \bibnamefont {Niu}}, \bibinfo {author} {\bibfnamefont
  {Y.}~\bibnamefont {Song}}, \bibinfo {author} {\bibfnamefont {C.~L.}\
  \bibnamefont {Zhang}}, \bibinfo {author} {\bibfnamefont {P.~C.}\ \bibnamefont
  {Dai}}, \bibinfo {author} {\bibfnamefont {B.~P.}\ \bibnamefont {Xie}},
  \bibinfo {author} {\bibfnamefont {X.~C.}\ \bibnamefont {Lai}}, \ and\
  \bibinfo {author} {\bibfnamefont {D.~L.}\ \bibnamefont {Feng}},\ }\href@noop
  {} {\bibfield  {journal} {\bibinfo  {journal} {Nat. Sci. Rep.}\ }\textbf
  {\bibinfo {volume} {5}},\ \bibinfo {pages} {9515} (\bibinfo {year}
  {2015})}\BibitemShut {NoStop}%
\bibitem [{\citenamefont {Song}\ \emph {et~al.}(2016)\citenamefont {Song},
  \citenamefont {Yan}, \citenamefont {Ye}, \citenamefont {Ren}, \citenamefont
  {Xu}, \citenamefont {Tan}, \citenamefont {Niu}, \citenamefont {Xie},
  \citenamefont {Zhang}, \citenamefont {Peng}, \citenamefont {Xu},
  \citenamefont {Jiang},\ and\ \citenamefont {Feng}}]{Song2016}%
  \BibitemOpen
  \bibfield  {author} {\bibinfo {author} {\bibfnamefont {Q.}~\bibnamefont
  {Song}}, \bibinfo {author} {\bibfnamefont {Y.~J.}\ \bibnamefont {Yan}},
  \bibinfo {author} {\bibfnamefont {Z.~R.}\ \bibnamefont {Ye}}, \bibinfo
  {author} {\bibfnamefont {M.~Q.}\ \bibnamefont {Ren}}, \bibinfo {author}
  {\bibfnamefont {D.~F.}\ \bibnamefont {Xu}}, \bibinfo {author} {\bibfnamefont
  {S.~Y.}\ \bibnamefont {Tan}}, \bibinfo {author} {\bibfnamefont {X.~H.}\
  \bibnamefont {Niu}}, \bibinfo {author} {\bibfnamefont {B.~P.}\ \bibnamefont
  {Xie}}, \bibinfo {author} {\bibfnamefont {T.}~\bibnamefont {Zhang}}, \bibinfo
  {author} {\bibfnamefont {R.}~\bibnamefont {Peng}}, \bibinfo {author}
  {\bibfnamefont {H.~C.}\ \bibnamefont {Xu}}, \bibinfo {author} {\bibfnamefont
  {J.}~\bibnamefont {Jiang}}, \ and\ \bibinfo {author} {\bibfnamefont {D.~L.}\
  \bibnamefont {Feng}},\ }\href@noop {} {\bibfield  {journal} {\bibinfo
  {journal} {Phys. Rev. B}\ }\textbf {\bibinfo {volume} {93}},\ \bibinfo
  {pages} {024508} (\bibinfo {year} {2016})}\BibitemShut {NoStop}%
\bibitem [{\citenamefont {Nozaki}\ \emph {et~al.}(2013)\citenamefont {Nozaki},
  \citenamefont {Nakano}, \citenamefont {Yajima}, \citenamefont {Kageyama},
  \citenamefont {Frandsen}, \citenamefont {Liu}, \citenamefont {Cheung},
  \citenamefont {Goko}, \citenamefont {Uemura}, \citenamefont {Munsie},
  \citenamefont {Medina}, \citenamefont {Luke}, \citenamefont {Munevar},
  \citenamefont {Nishio-Hamane},\ and\ \citenamefont {Brown}}]{Nozaki2013}%
  \BibitemOpen
  \bibfield  {author} {\bibinfo {author} {\bibfnamefont {Y.}~\bibnamefont
  {Nozaki}}, \bibinfo {author} {\bibfnamefont {K.}~\bibnamefont {Nakano}},
  \bibinfo {author} {\bibfnamefont {T.}~\bibnamefont {Yajima}}, \bibinfo
  {author} {\bibfnamefont {H.}~\bibnamefont {Kageyama}}, \bibinfo {author}
  {\bibfnamefont {B.}~\bibnamefont {Frandsen}}, \bibinfo {author}
  {\bibfnamefont {L.}~\bibnamefont {Liu}}, \bibinfo {author} {\bibfnamefont
  {S.}~\bibnamefont {Cheung}}, \bibinfo {author} {\bibfnamefont
  {T.}~\bibnamefont {Goko}}, \bibinfo {author} {\bibfnamefont {Y.~J.}\
  \bibnamefont {Uemura}}, \bibinfo {author} {\bibfnamefont {T.~S.~J.}\
  \bibnamefont {Munsie}}, \bibinfo {author} {\bibfnamefont {T.}~\bibnamefont
  {Medina}}, \bibinfo {author} {\bibfnamefont {G.~M.}\ \bibnamefont {Luke}},
  \bibinfo {author} {\bibfnamefont {J.}~\bibnamefont {Munevar}}, \bibinfo
  {author} {\bibfnamefont {D.}~\bibnamefont {Nishio-Hamane}}, \ and\ \bibinfo
  {author} {\bibfnamefont {C.~M.}\ \bibnamefont {Brown}},\ }\href {\doibase
  10.1103/PhysRevB.88.214506} {\bibfield  {journal} {\bibinfo  {journal} {Phys.
  Rev. B}\ }\textbf {\bibinfo {volume} {88}},\ \bibinfo {pages} {214506}
  (\bibinfo {year} {2013})}\BibitemShut {NoStop}%
\bibitem [{\citenamefont {Frandsen}\ \emph {et~al.}(2014)\citenamefont
  {Frandsen}, \citenamefont {Bozin}, \citenamefont {Hu}, \citenamefont {Zhu},
  \citenamefont {Nozaki}, \citenamefont {Kageyama}, \citenamefont {Uemura},
  \citenamefont {Yin},\ and\ \citenamefont {Billinge}}]{Frandsen2014}%
  \BibitemOpen
  \bibfield  {author} {\bibinfo {author} {\bibfnamefont {B.~A.}\ \bibnamefont
  {Frandsen}}, \bibinfo {author} {\bibfnamefont {E.~S.}\ \bibnamefont {Bozin}},
  \bibinfo {author} {\bibfnamefont {H.}~\bibnamefont {Hu}}, \bibinfo {author}
  {\bibfnamefont {Y.}~\bibnamefont {Zhu}}, \bibinfo {author} {\bibfnamefont
  {Y.}~\bibnamefont {Nozaki}}, \bibinfo {author} {\bibfnamefont
  {H.}~\bibnamefont {Kageyama}}, \bibinfo {author} {\bibfnamefont {Y.~J.}\
  \bibnamefont {Uemura}}, \bibinfo {author} {\bibfnamefont {W.-G.}\
  \bibnamefont {Yin}}, \ and\ \bibinfo {author} {\bibfnamefont {S.~J.~L.}\
  \bibnamefont {Billinge}},\ }\href {http://dx.doi.org/10.1038/ncomms6761}
  {\bibfield  {journal} {\bibinfo  {journal} {Nat. Commun.}\ }\textbf {\bibinfo
  {volume} {5}},\ \bibinfo {pages} {5761} (\bibinfo {year} {2014})}\BibitemShut
  {NoStop}%
\bibitem [{\citenamefont {Kitagawa}\ \emph {et~al.}(2013)\citenamefont
  {Kitagawa}, \citenamefont {Ishida}, \citenamefont {Nakano}, \citenamefont
  {Yajima},\ and\ \citenamefont {Kageyama}}]{Kitagawa2013}%
  \BibitemOpen
  \bibfield  {author} {\bibinfo {author} {\bibfnamefont {S.}~\bibnamefont
  {Kitagawa}}, \bibinfo {author} {\bibfnamefont {K.}~\bibnamefont {Ishida}},
  \bibinfo {author} {\bibfnamefont {K.}~\bibnamefont {Nakano}}, \bibinfo
  {author} {\bibfnamefont {T.}~\bibnamefont {Yajima}}, \ and\ \bibinfo {author}
  {\bibfnamefont {H.}~\bibnamefont {Kageyama}},\ }\href {\doibase
  10.1103/PhysRevB.87.060510} {\bibfield  {journal} {\bibinfo  {journal} {Phys.
  Rev. B}\ }\textbf {\bibinfo {volume} {87}},\ \bibinfo {pages} {060510}
  (\bibinfo {year} {2013})}\BibitemShut {NoStop}%
\bibitem [{\citenamefont {Chen}\ \emph {et~al.}(2016)\citenamefont {Chen},
  \citenamefont {Zhang}, \citenamefont {Song}, \citenamefont {Li},
  \citenamefont {Zhang}, \citenamefont {Qian}, \citenamefont {Luo},
  \citenamefont {Shi}, \citenamefont {Fang}, \citenamefont {Richard},\ and\
  \citenamefont {Ding}}]{Chen2016}%
  \BibitemOpen
  \bibfield  {author} {\bibinfo {author} {\bibfnamefont {D.}~\bibnamefont
  {Chen}}, \bibinfo {author} {\bibfnamefont {T.-T.}\ \bibnamefont {Zhang}},
  \bibinfo {author} {\bibfnamefont {Z.-D.}\ \bibnamefont {Song}}, \bibinfo
  {author} {\bibfnamefont {H.}~\bibnamefont {Li}}, \bibinfo {author}
  {\bibfnamefont {W.-L.}\ \bibnamefont {Zhang}}, \bibinfo {author}
  {\bibfnamefont {T.}~\bibnamefont {Qian}}, \bibinfo {author} {\bibfnamefont
  {J.-L.}\ \bibnamefont {Luo}}, \bibinfo {author} {\bibfnamefont {Y.-G.}\
  \bibnamefont {Shi}}, \bibinfo {author} {\bibfnamefont {Z.}~\bibnamefont
  {Fang}}, \bibinfo {author} {\bibfnamefont {P.}~\bibnamefont {Richard}}, \
  and\ \bibinfo {author} {\bibfnamefont {H.}~\bibnamefont {Ding}},\ }\href
  {\doibase 10.1103/PhysRevB.93.140501} {\bibfield  {journal} {\bibinfo
  {journal} {Phys. Rev. B}\ }\textbf {\bibinfo {volume} {93}},\ \bibinfo
  {pages} {140501} (\bibinfo {year} {2016})}\BibitemShut {NoStop}%
\bibitem [{\citenamefont {von Rohr}\ \emph {et~al.}(2013)\citenamefont {von
  Rohr}, \citenamefont {Schilling}, \citenamefont {Nesper}, \citenamefont
  {Baines},\ and\ \citenamefont {Bendele}}]{vonRohr2013}%
  \BibitemOpen
  \bibfield  {author} {\bibinfo {author} {\bibfnamefont {F.}~\bibnamefont {von
  Rohr}}, \bibinfo {author} {\bibfnamefont {A.}~\bibnamefont {Schilling}},
  \bibinfo {author} {\bibfnamefont {R.}~\bibnamefont {Nesper}}, \bibinfo
  {author} {\bibfnamefont {C.}~\bibnamefont {Baines}}, \ and\ \bibinfo {author}
  {\bibfnamefont {M.}~\bibnamefont {Bendele}},\ }\href {\doibase
  10.1103/PhysRevB.88.140501} {\bibfield  {journal} {\bibinfo  {journal} {Phys.
  Rev. B}\ }\textbf {\bibinfo {volume} {88}},\ \bibinfo {pages} {140501(R)}
  (\bibinfo {year} {2013})}\BibitemShut {NoStop}%
\bibitem [{\citenamefont {Nakaoka}\ \emph {et~al.}(2016)\citenamefont
  {Nakaoka}, \citenamefont {Yamakawa},\ and\ \citenamefont
  {Kontani}}]{Nakaoka2016}%
  \BibitemOpen
  \bibfield  {author} {\bibinfo {author} {\bibfnamefont {H.}~\bibnamefont
  {Nakaoka}}, \bibinfo {author} {\bibfnamefont {Y.}~\bibnamefont {Yamakawa}}, \
  and\ \bibinfo {author} {\bibfnamefont {H.}~\bibnamefont {Kontani}},\ }\href
  {\doibase 10.1103/PhysRevB.93.245122} {\bibfield  {journal} {\bibinfo
  {journal} {Phys. Rev. B}\ }\textbf {\bibinfo {volume} {93}},\ \bibinfo
  {pages} {245122} (\bibinfo {year} {2016})}\BibitemShut {NoStop}%
\bibitem [{\citenamefont {Kim}\ and\ \citenamefont {Kee}(2015)}]{Kim2015}%
  \BibitemOpen
  \bibfield  {author} {\bibinfo {author} {\bibfnamefont {H.~S.}\ \bibnamefont
  {Kim}}\ and\ \bibinfo {author} {\bibfnamefont {H.~Y.}\ \bibnamefont {Kee}},\
  }\href {\doibase 10.1103/PhysRevB.92.235121} {\bibfield  {journal} {\bibinfo
  {journal} {Phys. Rev. B}\ }\textbf {\bibinfo {volume} {92}},\ \bibinfo
  {pages} {235121} (\bibinfo {year} {2015})}\BibitemShut {NoStop}%
\bibitem [{\citenamefont {Shi}\ \emph {et~al.}(2013)\citenamefont {Shi},
  \citenamefont {Wang}, \citenamefont {Zhang}, \citenamefont {Wang},
  \citenamefont {Huang},\ and\ \citenamefont {Wang}}]{Shi2013}%
  \BibitemOpen
  \bibfield  {author} {\bibinfo {author} {\bibfnamefont {Y.~G.}\ \bibnamefont
  {Shi}}, \bibinfo {author} {\bibfnamefont {H.~P.}\ \bibnamefont {Wang}},
  \bibinfo {author} {\bibfnamefont {X.}~\bibnamefont {Zhang}}, \bibinfo
  {author} {\bibfnamefont {W.~D.}\ \bibnamefont {Wang}}, \bibinfo {author}
  {\bibfnamefont {Y.}~\bibnamefont {Huang}}, \ and\ \bibinfo {author}
  {\bibfnamefont {N.~L.}\ \bibnamefont {Wang}},\ }\href {\doibase
  10.1103/PhysRevB.88.144513} {\bibfield  {journal} {\bibinfo  {journal} {Phys.
  Rev. B}\ }\textbf {\bibinfo {volume} {88}},\ \bibinfo {pages} {144513}
  (\bibinfo {year} {2013})}\BibitemShut {NoStop}%
\bibitem [{\citenamefont {Nowell}\ \emph {et~al.}(2012)\citenamefont {Nowell},
  \citenamefont {Barnett}, \citenamefont {Christensen}, \citenamefont {Teat},\
  and\ \citenamefont {Allan}}]{Nowell2012}%
  \BibitemOpen
  \bibfield  {author} {\bibinfo {author} {\bibfnamefont {H.}~\bibnamefont
  {Nowell}}, \bibinfo {author} {\bibfnamefont {S.~A.}\ \bibnamefont {Barnett}},
  \bibinfo {author} {\bibfnamefont {K.~E.}\ \bibnamefont {Christensen}},
  \bibinfo {author} {\bibfnamefont {S.~J.}\ \bibnamefont {Teat}}, \ and\
  \bibinfo {author} {\bibfnamefont {D.~R.}\ \bibnamefont {Allan}},\ }\href
  {\doibase 10.1107/S0909049512008801} {\bibfield  {journal} {\bibinfo
  {journal} {Journal of Synchrotron Radiation}\ }\textbf {\bibinfo {volume}
  {19}},\ \bibinfo {pages} {435} (\bibinfo {year} {2012})}\BibitemShut
  {NoStop}%
\bibitem [{\citenamefont {Campbell}\ \emph {et~al.}(2006)\citenamefont
  {Campbell}, \citenamefont {Stokes}, \citenamefont {Tanner},\ and\
  \citenamefont {Hatch}}]{Campbell2006}%
  \BibitemOpen
  \bibfield  {author} {\bibinfo {author} {\bibfnamefont {B.~J.}\ \bibnamefont
  {Campbell}}, \bibinfo {author} {\bibfnamefont {H.~T.}\ \bibnamefont
  {Stokes}}, \bibinfo {author} {\bibfnamefont {D.~E.}\ \bibnamefont {Tanner}},
  \ and\ \bibinfo {author} {\bibfnamefont {D.~M.}\ \bibnamefont {Hatch}},\
  }\href {\doibase http://dx.doi.org/10.1107/S0021889806014075} {\bibfield
  {journal} {\bibinfo  {journal} {J. Appl. Cryst.}\ }\textbf {\bibinfo {volume}
  {39}},\ \bibinfo {pages} {607 } (\bibinfo {year} {2006})}\BibitemShut
  {NoStop}%
\bibitem [{\citenamefont {Miller}\ and\ \citenamefont
  {Love}(1967)}]{Miller1967}%
  \BibitemOpen
  \bibfield  {author} {\bibinfo {author} {\bibfnamefont {S.~C.}\ \bibnamefont
  {Miller}}\ and\ \bibinfo {author} {\bibfnamefont {W.~F.}\ \bibnamefont
  {Love}},\ }\href@noop {} {\emph {\bibinfo {title} {Tables of Irreducible
  Representations of Space Groups and Co-Representations of Magnetic Space
  Groups}}}\ (\bibinfo  {publisher} {Pruett Press},\ \bibinfo {year}
  {1967})\BibitemShut {NoStop}%
\bibitem [{\citenamefont {Rodriguez-Carvajal}(1993)}]{RodriguezCarvajal1993}%
  \BibitemOpen
  \bibfield  {author} {\bibinfo {author} {\bibfnamefont {J.}~\bibnamefont
  {Rodriguez-Carvajal}},\ }\href {\doibase
  http://dx.doi.org/10.1016/0921-4526(93)90108-I} {\bibfield  {journal}
  {\bibinfo  {journal} {Physica B: Condensed Matter}\ }\textbf {\bibinfo
  {volume} {192}},\ \bibinfo {pages} {55} (\bibinfo {year} {1993})}\BibitemShut
  {NoStop}%
\end{thebibliography}%

\newpage

\section*{Appendix}

\subsection*{Symmetry Analysis and Fitting}

\begin{figure*}[t]
	\centering
	\includegraphics[width=0.9\textwidth]{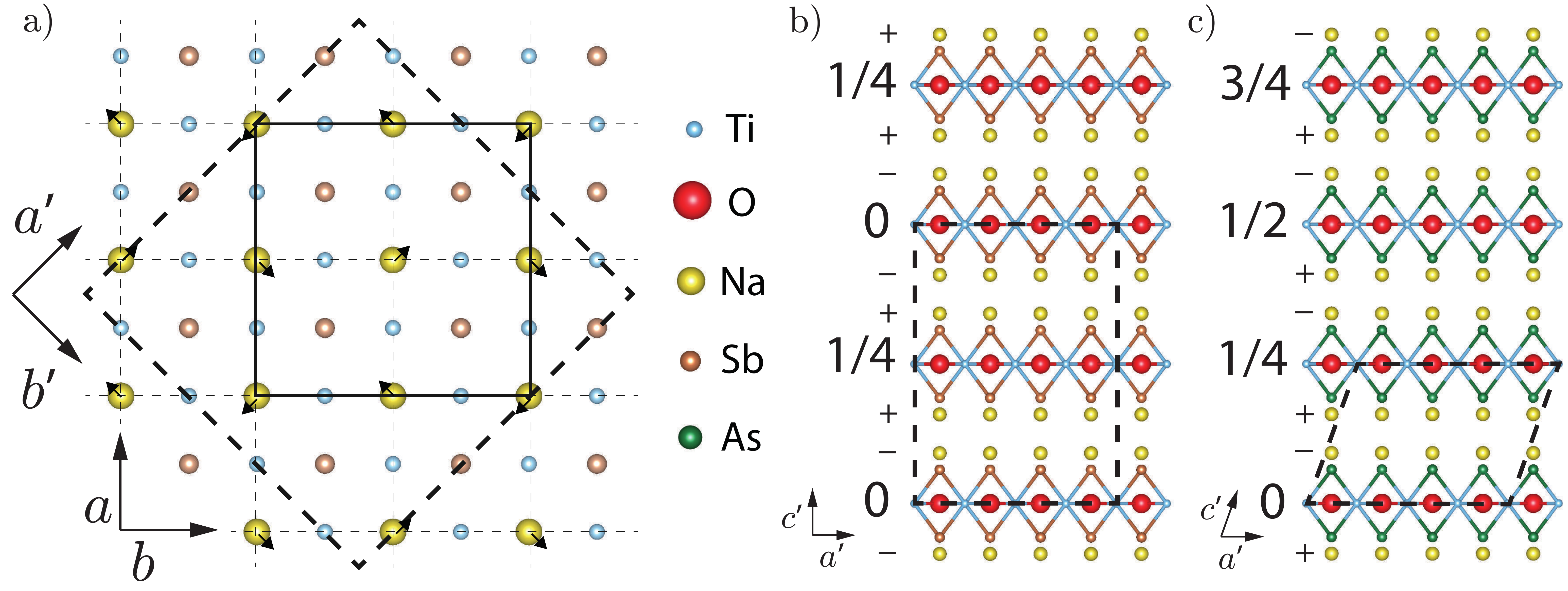}
	\caption{The proposed distortion in the Na layers. (a) $ab$-plane distortion pattern for a single Na layer relative to the positions of the Ti and Sb atoms in the nearest Ti$_2Pn_2$O unit, with thin dashed lines showing the high temperature $I4/mmm$ unit cell boundaries, thick dashed lines the low temperature unit cell boundary and solid lines the unit cell of the Ti$_2Pn_2$O layer distortion pattern. (b) and (c) show how these layers stack, with $+$ and $-$ beside each Na layer being the sign of the distortion in that layer, where $+$ is defined as the pattern in (a), and the numbers beside each Ti$_2Pn_2$O unit giving the distortion pattern in that unit as defined in the main text. Black dashed lines mark the unit cell of the distorted structure in each case.}
	\label{Xtal_Struct_Fig_SI}
\end{figure*}

\begin{table*}[]
	\centering
	\begin{tabular}{c|cccc||c|ccc}
		$Pn=$Sb & & & &&$Pn=$As & & & \\
		$Cmcm$& $x$ & $y$ & $z$ &&$C2/m$& $x$ & $y$ & $z$ \\
		\hline
		Na1& 0 & 0.377(3) & 0.93054&& Na1& 0.9097 & 0 & 0.6389 \\
		Na2& 0 & 0.127(3) & 0.43054 &&Na2& 0.4097 & 0 & 0.6389 \\
		Na3& 0.752(3) & 0.125 & 0.93054 &&Na3& 0.1597 & 0.25 & 0.6389 \\
		Ti1& 0.3668(12) & 0.0082(12) & 0.25 &&Ti1& 0.1315(14) & 0.1185(14) & 0 \\
		Ti2& 0.6332(12) & 0.2418(12) & 0.25 &&Ti2& 0.8815(14) & 0.3685(14) & 0 \\
		Ti3& 0.8832(12) & -0.0082(12) & 0.25 &&As1& 0.810(1) & 0 & 0.2423 \\
		Ti4& 0.1168(12) & 0.2582(12) & 0.25 &&As2& 0.311(1) & 0 & 0.2423 \\
		Sb1& 0 & 0.376(1) & 0.12884 &&As3& 0.060(1) & 0 & 0.2423 \\
		Sb2& 0 & 0.126(1) & 0.62884 &&O1& 0 & 0 & 0 \\
		Sb3& 0.751(1) & 0.125 & 0.12884 &&O2& 0 & 0.5 & 0 \\
		O1& 0 & 0.625 & 0.25 &&O3& 0.25 & 0.25 & 0 \\
		O2& 0 & 0.125 & 0.25 &&&&&\\
		O3& 0.75 & 0.375 & 0.25 &&&&&\\
	\end{tabular}
	\caption{Atomic positions in the distorted structure of Na$_{2}$Ti$_{2}Pn_{2}$O for $Pn$ = Sb and As from the least-squares fits to our X-ray data shown in the main text.}
	\label{Atomic_Pos_Table}
\end{table*}

For $Pn =$ Sb, all reflections from the model distorted structure were found to obey $F_{hkl} = \pm F_{khl}$ where $hkl$ are referred to the reciprocal space of the $I4/mmm$ structure, so that the contributions to intensity at any $hkl$ from the two equivalent orthorhombic domains are equal. It was therefore possible to perform a quantitative refinement by using a single domain and refining an overall scale factor. For $Pn =$ As, it was found that there is no overlap between superstructure reflections from the two equivalent domains so a quantitative fit was again possible by indexing half of the observed superstructure peaks using a single domain.

Displacive distortion modes of the Na$_{2}$Ti$_{2}$Sb$_{2}$O $I4/mmm$ parent crystal structure were calculated using Isodistort.\cite{Campbell2006} Four irreducible representations of the commensurate propagation vector $(1/2, 0, 0)$ were determined, labelled $\Sigma 1$, $\Sigma 2$, $\Sigma 3$, and $\Sigma 4$ in Miller and Love notation,\cite{Miller1967} which support finite displacements of all atomic species present. Considering all symmetry-distinct directions of the order parameter, a total of 52 superstructures were found. As discussed in the main text, the structural distortion model was constrained to include exactly two active propagation vectors, ${\bf q}_{1}$ and ${\bf q}_2$, reducing the total number of possible superstructures to 40. To constrain the number of models further, we considered two qualitative features of the experimental diffraction pattern. Firstly, reflections extinct by $I$-centering in the high temperature parent phase at $h+k+l = $ odd positions were observed to remain extinct in the distorted phase. The super-space group must therefore contain a symmetry element that, for all atomic positions, gives a reflection condition equivalent to that of the $I$-centering translational symmetry of the parent. This strict constraint limits the model to one of eight superstructures. Secondly, the period-eight modulation observed in the diffraction intensity along $l$ requires displacements of both Ti and Sb ions to be orthogonal to the $I4/mmm$ $c$-axis, giving just two possible superstructures. They are both described by the space group $Cmcm$ with basis $\{(2,2,0),(2,-2,0),(0,0,-1)\}$ and origin shift $(3/4,5/4,-1/4)$ with respect to the parent structure, and within $\Sigma 1$ and $\Sigma 4$ representations, respectively. The $\Sigma 1$ mode describes displacements of the titanium ions parallel to the respective O--Ti--O bond, whereas the $\Sigma 4$ mode describes orthogonal displacements.

Refinement of both $\Sigma 1$ and $\Sigma 4$ $Cmcm$ distortion modes against the diffraction data using the FullProf software package \cite{RodriguezCarvajal1993} shows that both models are in reasonable qualitative agreement with the data, but only the $\Sigma 4$ model accurately reproduces the relative intensity between groups of reflections along $l$, when measured at different $h$ or $k$ as seen in Fig.~2 in the main text.

The structural distortion model for Na$_{2}$Ti$_{2}$As$_{2}$O was found via the same strategy.

For both materials, displacements of the Ti and Sb/As sites are required to reproduce all of the qualitative features of the experimental data. A small displacement of the Na sites is also allowed by symmetry and was found to improve the quantitative fit slightly for $Pn =$ Sb, although the fit is not very sensitive to this. Such a displacement would also be allowed for $Pn =$ As, although our data are not of sufficient quality to perform a quantitative refinement in that case. The relevant displacement pattern for Na layers and how the layers stack along $c$ is presented in Figure~\ref{Xtal_Struct_Fig_SI}, while final positions for all sites in the best-fit distorted structures can be found in Table~\ref{Atomic_Pos_Table}.

\end{document}